\documentclass[a4, 12pt]{article}
\usepackage[linesnumbered,ruled,vlined]{algorithm2e}
\usepackage[hyphenbreaks]{breakurl}
\usepackage[justification=centering]{caption}
\usepackage{wrapfig, color}
\usepackage{dblfloatfix}    
\usepackage[margin=1in]{geometry}
\usepackage{diagbox,threeparttable} 
\usepackage[sort, round]{natbib}
\usepackage{amsmath,amsthm, amssymb,enumitem,esvect,bbm}
\usepackage{graphicx, wrapfig}
\usepackage{comment, setspace, empheq}
\usepackage[T1]{fontenc}
\usepackage[breaklinks]{hyperref}
\usepackage{textcomp, multirow}
\usepackage{algorithmic, url}

\allowdisplaybreaks
\usepackage[compact]{titlesec}  
\titlespacing*{\section}{0pt}{0pt}{0pt}
\titlespacing*{\subsection}{0pt}{0pt}{0pt}

\def\thm@space@setup{%
  \thm@preskip=0pt plus 1pt minus 1pt
  \thm@postskip=\thm@preskip 
}
\newtheorem{defn}{Definition}

\newcommand{\GI}{geo-indistinguishability}

\newcommand{\D}{doppelganger}
\newcommand{\bs}{\boldsymbol}
\newcommand{\M}{\mathcal{M}}
\def\BibTeX{{\mathcal{M}m B\kern-.05em{\sc i\kern-.025em b}\kern-.08em
    T\kern-.1667em\lower.7ex\hbox{E}\kern-.125emX}}


\begin{document}
\title{\LARGE{Some Examples of Privacy-preserving Publication and Sharing of COVID-19 Pandemic Data}\vspace{-0.5em}}
\author{Fang Liu$^1$\footnote{Corresponding author: Fang Liu (fang.Liu.131@nd.edu).  Fang Liu was supported by NSF Grant NO. 1717417 and  the University of Notre Dame Asia Research Collaboration Grant.  Dong Wang was supported by the  China Scholarships Council program (NO. 201906270230) and NSFC Grant NO. 41971407; Tian Yan was supported by the University of Notre Dame Asia Research Collaboration Grant and the Chinese Scholarship Council,}, Dong Wang$^2$, Tian Yan$^1$\\
$^1$\small{Department of Applied and Computational Mathematics and Statistics}\\
\small{University of Notre Dame, IN 46556, USA}\\
$^2$\small{College of Cyberspace Security} \\
\small{Hangzhou Dianzi University, Wuhan, 430079, China}
\vspace{-3em}}
\date{}
\maketitle
\begin{abstract}
A considerable amount of various types of data have been collected during the COVID-19 pandemic, the analysis and interpretation of which have been indispensable for curbing the spread of the disease. As the pandemic moves to an endemic state, the data collected during the pandemic will continue to be rich sources for further studying and understanding the impacts of the pandemic on various aspects of our society.  On the other hand, na\"{i}ve release and sharing of the information can be associated with serious privacy concerns.  In this study, we use three common but distinct data types collected during the pandemic (case surveillance tabular data, case location data, and contact tracing networks) to illustrate the publication and sharing of granular information and individual-level pandemic data in a  privacy-preserving manner. We leverage and build upon the concept of differential privacy to generate and release privacy-preserving data for each data type. We investigate the inferential utility of privacy-preserving information through simulation studies at different levels of privacy guarantees and demonstrate the approaches in real-life data.  All the approaches employed in the study are straightforward to apply. Our study generates statistical evidence on the practical feasibility of sharing pandemic data with privacy guarantees and on how to balance the statistical utility of released information during this process. 

\vspace{6pt}
\noindent \textbf{keywords}: COVID-19  pandemic,  differential privacy,  \GI,  hot spot heat maps, contact tracing network, synthetic data
\end{abstract} 

\flushbottom
\maketitle

\setlength{\abovedisplayskip}{6pt}
\setlength{\belowdisplayskip}{6pt}  

\section{Introduction}\label{sec:introduction}
A huge amount of data of various types have been collected during the  COVID-19  pandemic, the analysis and interpretation of which has been indispensable to  health authorities and experts to gain understanding of the disease, to identify risk factors, to monitor and forecast the spread of the disease, to evaluate the impacts of the pandemic on different aspects of our society,  and to implement strategies that mitigates negative impacts. As the pandemic shifts to an endemic state, the collected data will continue to serve as rich sources for further research on the disease and its impacts so to prepare us for  future pandemics.

Na\"{i}ve release and sharing of the pandemic data can be associated with serious privacy concerns, especially considering that a huge amount and a great variety of data were collected quickly in a short period of time and the data privacy and ethics regulations were lagging behind at least in the initial stage of the pandemic. Many types of collected data are known to be associated with high privacy risk, such as disease status, medical history, locations, close contacts, employment/income status, etc.  Balance must be found between individual privacy protection and sharing the data for research use.  

Fortunately, this is not an unsolvable problem as many research questions of interest, the answers to which lie within the pandemic data, revolve around learning population-level and aggregate information rather than focusing on individual-level information which is often the goal of privacy attacks. If a privacy-preserving data release procedure can maintain accurate and useful aggregate information while guaranteeing individual-level privacy, it would make a potentially effective approach for data sharing. 

Privacy-preserving collection and analysis of COVID-19 data have been developed and applied during the pandemic. Google research teams apply differential privacy (DP) to  generate anonymized metrics from the data of Google users who opted in for the Location History setting in their Google accounts and produce the COVID-19 community mobility reports \citep{aktay2020google},  to understand the impacts of social distancing policies on mobility and COVID-19 case growth in the US \citep{wellenius2021impacts}, to generate anonymized trends in Google searches for COVID-19 symptoms and related topics \citep{fabrikant2020google}, and to forecast COVID-19 trends using spatio-temporal graph neural networks \citep{kapoor2020examining}. DP is also integrated in deep learning to predict COVID-19 infections from imaging data \citep{muftuouglu2020differential, ulhaq2020COVID}. \citet{butler2020differentially} apply DP to generate individual-level health tokens/randomized health certificates while allowing useful aggregate risk estimates to be calculated.   The methods in all the work above are designed to provide the usefulness of collective information in the released data without disclosing individual-level information upon release.

Location data and proximity data has been instrumental to track the trajectory of a COVID-19 case and for contact tracing (CT) so to identify people who might have close contacts with COVID-19 patients. On the other hand,  location and relational information can be highly revealing of personal information in general. 
To protect the sensitive information, privacy-preserving technologies and tools are  adopted in  CT apps and software around the world during the pandemic to track the spread of the disease. The apps collect users' location data (e.g., GPS) or proximity data (e.g., Bluetooth), via either a centralized (e.g., Alipay Health Code and WeChat in China \citep{china2},  Corona100m in South Korea \citep{southkorea}, COVIDTracker in Thailand  \citep{Thailand}, ProteGo in Poland \citep{Poland}, and Pan-European Privacy-Preserving Proximity Tracing (PEPP-PT) in EU \citep{EU}) or decentralized model (Safe paths  \citep{safepath}  and the proximity-based Google/Apple Exposure Notification (GAEN) system  \citep{gaen} in the US) to identify and notify those who might have been near a COVID-19 patient and at high risk of contracting the disease. We refer readers to \citet{wang2020privacy} for a compensate review of the CT apps used during the pandemic.    

Many privacy-preserving methods developed and implemented during the pandemic, including the work reviewed above, focus on information shared  with governments, health officials, and the public so to facilitate  quick decision making and timely actions during the pandemic. In contrast, privacy-preserving COVID-19 data release for research use has received less attention, which is the major focus of our work. Sharing data for research use is not just for making scientific discoveries, but also for producing real-world evidence and generating new insights into how we can better handle similar crisis in the future. Data for research use often contain granular information compared to those shared with administrators,  decision makers, and the public, and thus are associated with higher privacy risk that must be mitigated before release.  

We leverage and build upon existing DP concepts  and techniques  and apply them to several common but distinct  pandemic data types to publish data with formal privacy guarantees. 
It is not our goal to cover all types of data collected during the pandemic,  but rather use three  pandemic data types -- surveillance data, case location data, and contact tracing networks (CTNs)  -- to demonstrate how to apply formal privacy concepts to release privacy-preserving information. 
We choose the three data types because they  data routinely collected during the pandemic, they  are distinct in terms of data structure and statistical analysis, and they provide different information on COVID-19. Specifically, surveillance data help better understand risk factors associated with COVID-19 and identify sub-populations that are vulnerable to the disease; location data can be used to explore relationships between hot spots and residential characteristics to study issues such as residential racism and structural segregation during  the pandemic, CTNs allow us to study how clustering of COVID-19 cases and how physical proximity may affect the spread of the disease, among others.

In all three data types, we focus on release of synthetic data generated at a pre-specified privacy budget. With synthetic data, data users may perform analysis on their own \citep{bowen2020comparative}. For surveillance data, we use the flat Laplace sanitizer with DP guarantees and examine  statistical  utility of log-linear models based on sanitized data, as a function of sample size and pre-specified privacy budget in simulation studies and real data published by the U.S. CDC. For location data,  we apply the planar Laplace mechanism with geo-indistinguishability guarantees to release data. We conduct simulation studies and apply the method to  a real South Korean case location dataset to examine inference from  cluster point process models and accuracy of hot spot heat maps based on  sanitized locations. For CTN data,   we examine the feasibility of the approach of differentially private exponential random graph models (ERGMs) to generate privacy-preserving synthetic networks. We conduct simulation studies to investigate the  utility of sanitized networks in inference from  ERGMs and  preservation of descriptive structural network information. 


The rest of the paper is organized as follows. Section \ref{sec:prelim} provides an overview of the basic concepts in DP, some common randomized mechanisms for achieving DP, and an approach for obtaining valid inferences from sanitized data. Section \ref{sec:surveillance}, \ref{sec:location} and \ref{sec:ct} apply DP procedures to release privacy-preserving case surveillance data, case location data, and CTNs, respectively, conduct simulation studies to examine the statistical utility of the privacy-preserving data, and apply the DP procedures to real pandemic data. Section \ref{sec:discussion} provides some final remarks on the implementations of DP methods in releasing COVID-19 data.

\section{Preliminaries}\label{sec:prelim}\vspace{-3pt}
We provide a brief overview of some common DP concepts and mechanisms. The  overview is not comprehensive and  does not aim at covering every  concept in DP, but rather focuses on those  used or mentioned in this paper. 

\subsection{Differential privacy}
\begin{defn}[$(\epsilon,\delta)$-DP\citep{dwork2006calibrating, dwork2006our}] \label{defn:dp}
A randomized algorithm $\M$ is of $(\epsilon,\delta)$-DP if for  all dataset pairs of neighboring data sets $(D,D')$ differing by one record and for all subsets $\mathcal{S}\subseteq$ image$(\M)$,
\begin{equation}\label{eqn:DP}
\Pr(\mathcal{M}(D)\in \mathcal{S}) \leq e^{\epsilon} \Pr(\mathcal{M}(D')\in \mathcal{S})+\delta.
\end{equation}
\end{defn}\vspace{-6pt}
$D$ and $D'$ differing by one record (denoted by $d(D,D')=1$) may refer to the case that they are of the same size but differ in at least one  attribute value in exactly one record (bounded DP), or  $D'$ has one record less than $D$ or vice versa (unbounded DP) \citep{Kifer2011}.  $\epsilon>0$ and $\delta\ge0$ are privacy budget or privacy loss parameters.  When $\delta=0$, $(\epsilon,\delta)$-DP becomes  pure $\epsilon$-DP; the smaller $\epsilon$ is, the more privacy protection there is on any individual in the data, as the released results $\mathcal{M}(D)$ and $\mathcal{M}(D')$ are similar in the sense that their probability density/mass function ratio  is bounded with $(e^{-\epsilon}, e^{\epsilon})$.  There is no consensus and lacks a universal guideline on the choice of $\epsilon$ \citep{dwork2019differential}.  $\epsilon$ typically ranges from $10^{-3}$ to $10$ in empirical studies in the DP literature on, depending on the type of information released, social perception of privacy, expected accuracy of released data, among others. Real-life of applications of DP often employs larger $\epsilon$ for better utility (e.g., US Census uses $\epsilon$ of 19.61 \citep{census} and Apple Inc. sets $\epsilon$ at 2, 4, or 8  for different Apps \citep{apple}).  $\delta$, if not 0, is often set at a very small value (inverse proportional to poly$(n)$) and can be interpreted as the probability that the pure $\epsilon$-DP is violated.

Definition \ref{defn:dp} is the original DP definition. Relaxed versions and extensions exist, such as $(\epsilon,\delta)$-probabilistic DP (pDP) \citep{machanavajjhala2008privacy},  $(\epsilon,\tau)$-concentrated DP (CDP) \citep{cPD}, zero-concentrated DP \citep{bun2016concentrated} (zCDP), R\'{e}nyi DP (RDP) \citep{mironov2017renyi}, and Gaussian DP (GDP) \citep{dong2019gaussian}.

DP provides a mathematically rigorous framework for protecting individual privacy when releasing and sharing information. Many mechanisms and procedures have been developed to achieve DP. In this paper, we employ the Laplace mechanism with pure $\epsilon$-DP  to illustrate how to apply DP concepts and procedures to protect individual privacy when releasing COVID-19 data. When other types of DP guarantees are desired, such as $(\epsilon,\delta)$-(p)DP, corresponding mechanisms can be used, such as the Gaussian mechanisms \citep{dwork2014algorithmic,liu2018generalized}.

\begin{defn}[Laplace mechanism \citep{dwork2006calibrating}]\label{def:Laplace}  
Let $\mathbf{s}=(s_1,\ldots,s_r)$ be a statistic calculated from a dataset. The \textit{Laplace mechanism} of  $\epsilon$-DP releases $\mathbf{s}^*=\mathbf{s}+\mathbf{e}$, where $\mathbf{e}$ contains $r$ independent samples from Laplace$\left(0,\Delta\epsilon^{-1}\right)$, where $\Delta_1=\mbox{max}_{\scriptstyle{x,x', d(x,x')=1}} \|\mathbf{s}(x)-\mathbf{s}(x')\|_1$ is the $\ell_1$ \textit{global sensitivity} of $\mathbf{s}$. 
\end{defn}\vspace{-6pt}
The $\ell_1$ global sensitivity represents the maximum $\ell_1$ change  in $\mathbf{s}$ between two neighboring data sets (in general, one can define  $\ell_p$ $(p\ge0)$ global sensitivity; see \cite{liu2018generalized}). The larger the sensitivity, the more impact a single individual has on the value of  $\mathbf{s}$ and more noise would be needed to achieve $\epsilon$-DP.

Every time a dataset is queried, there is a privacy cost (loss) on the individuals in the dataset. Data curators need to track the privacy cost during the querying process to ensure the overall privacy spending does not exceed a pre-specified level.  Two basic composition principles in DP, \emph{parallel composition} and \emph{sequential composition} \citep{mcsherry2007mechanism}, can be used in privacy loss accounting, which are also used in later sections of the paper.
\begin{defn}[Basic privacy loss composition \citep{mcsherry2007mechanism}] \label{def:composition}  
If mechanism $\mathcal{M}_j$ of $(\epsilon_j,\delta_j)$-DP is applied to disjoint dataset $D_j$  for $j=1,\dots,P$, the parallel composition states the total privacy loss in data $\cup_j{D_j}$ from apply the $P$ mechanisms $\mathcal{M}_j$ for  $j=1,\dots,P$ is $(\max\{\epsilon_j\}, \max\{\delta_j\})$; if $\mathcal{M}_j$ is applied to the same dataset $D$, the sequential composition states that the total privacy loss in $D$ is $(\sum_j\epsilon_j, \sum_j\delta_j)$ from applying the $P$ mechanisms $\mathcal{M}_j$ for  $j=1,\dots,P$. 
\end{defn}\vspace{-9pt}
In layman's terms, the two privacy loss composition principle  states as long as there is no overlapping information between two datasets to which two DP mechanisms is applied to, the overall loss for releasing the queries results is the maximum privacy spending between the two; otherwise,  the loss adds up. The sequential composition on $(\epsilon,\delta)$-DP can be over-conservative for repeated querying on the same data; advanced composition \citep{dwork2010boosting} for $(\epsilon,\delta)$-DP and the relaxed DP notions mentioned above (e.g., CDP, zCDP, RDP, GDP) all achieve tighter total privacy loss bound than the basic composition. 

DP is a main-stream concept in privacy research and applications nowadays. Backed up  its mathematical rigor and robustness to various privacy attacks, the properties it has, including privacy loss composition, immunity to post-processing, and being future-proof,  make it attractive for designing sophisticated DP procedures and algorithms for complicated analysis and learning problems such as deep learning \citep{abadi2016deep} and regularized regressions \citep{chaudhuri2011differentially,kifer2012private,Li2022}. Immunity to post-processing and being future-proof refer to instances that information released from a DP mechanism won't leak additional information about the individuals in the dataset on which the information is based when it is further processed  after the release or when there is additional information on these individuals in the future from other sources, as long as  the original data is not accessed.

\subsection{Location privacy}
\citet{andres2013geo} extend the pure $\epsilon$-DP concept to releasing privacy-preserving location data that are  represented as pairs of 2-dimensional GPS coordinates, along with the planar Laplace mechanism to achieve such privacy guarantees.
\begin{defn}[Geo-indistinguishability (GI) \citep{andres2013geo}]label{def:GI}
Let $d(P,P')$  denote the Euclidean distance between any two distinct locations $P$ and $P'$, and $\epsilon$ be the unit-distance privacy loss. A randomized mechanism $\M$ satisfies $\epsilon$-GI if and only,  for any $\gamma>0$, any possible released location $P^*$, and all possible pairs of $P$ and $P'$ that $d(P,P')\le \gamma$,
\begin{equation}\label{Eq:geoDP}
\Pr(\M(P)=P^*|P)\le e^{\epsilon\gamma}\cdot\Pr(\M(P')=P^*|P').
\end{equation}
\end{defn}\vspace{-9pt}
$\M$ in Eq (\ref{Eq:geoDP}) enjoys $(\epsilon\gamma)$-GI for any specified $\gamma>0$ in the sense that the probability of distinguishing any two locations within a radius of $\gamma$, given the released location $P^*$, is $e^{\epsilon\gamma}$-fold the probability  when  not  having $P^*$. $\epsilon$ is the per-unit-distance loss and $\gamma$ denotes how many units. The larger $\epsilon$ is, the larger the privacy loss $(\epsilon\gamma)$ is and the higher probability of identifying the true location information within a radius of $\gamma$ mile given the perturbed location information. Though increasing $\gamma$ would also lead to higher privacy loss and probability identifying the true location is within a radius of $\gamma$  but the large $\gamma$ would make this identification less meaningful.

\begin{defn}[planar Laplace mechanism \citep{andres2013geo}]\label{def:polar}
Let the coordinates of the observed location $P$ in the Euclidean space be $(x,y)$. The planar Laplace mechanism of $\epsilon$-GI generates  sanitized location $P^*$ with coordinates
\begin{align}
&(x^*,y^*) =(x+r\cos(\theta), y+r\sin(\theta)), \mbox{ where} \label{Eq:xsys}\\
&r\sim\mbox{gamma }(2,\epsilon)=r\epsilon^2 e^{-\epsilon r}\mbox{ and }
\theta\sim\mbox{uniform }(0,2\pi)=1/(2\pi).\label{Eq:rtheta}
\end{align}
\end{defn}\vspace{-6pt}
$r$ in Eqs \eqref{Eq:rtheta} is the distance between $P^*$ and $P$  and $\theta$ is the angle of $P\rightarrow P^*$ in the Euclidean space, and  $r$ and $\theta$  are independent. The concepts of GI and planar Laplace mechanism are employed in Section \ref{sec:location} for releasing privacy-preserving location data.

More precisely speaking, GI is more related to local DP \citep{duchi2013local}, an extension of the pure $\epsilon$-DP, than the latter per se, which is often used for releasing aggregate information rather than individual response.
\begin{defn}[$\epsilon$-local DP \citep{duchi2013local}]\label{def:localDP} 
A randomization mechanism $\M$ provides $\epsilon$ local DP if 
$\Pr[\M(x)\in\Omega]\le{e^{\epsilon}}\cdot\Pr[\M(x')\in{\Omega}]$
for all pairs of possible data points $x$ and $x'$ from an individual and all possible output subset $\Omega$ from $\M$.
\end{defn}

\vspace{-3pt}\subsection{Privacy-preserving statistical inference} \label{sec:inference}\vspace{-3pt}
Sanitized outputs, compared to the original outputs, are subject to an extra source of variability due to the noise introduced through the randomized algorithm $\mathcal{R}$ for achieving DP. To account for the extra source of variability for valid statistical inference, one may directly model the sanitization mechanism, which may complicate the regular inferential procedures either analytically or computationally and is problem-specific. An alternative is the multiple synthesis (MS) approach that releases multiple sets of sanitized datasets or statistics and employs an inferential rule across the multiple sets to obtain valid inference \citet{liu2016model}. The MS approach is general and straightforward to apply. We adopt the MS approach to obtain privacy-preserving inference from sanitized data in this paper.

Denote the number of released sets by $m$. Per sequential composition, the total privacy budget would split into $m$ portions, one per release. $m\in[3,5]$ is recommended \citep{liu2016model}.  WLOS, suppose the parameter of interest is $\beta$ and its $l$-th sanitized estimate is $\hat{\beta}^{(l)}$  with estimated variance $w^{(l)}$ for $l=1,\ldots,m$.  The final inference of $\beta$, including hypothesis testing and confidence interval (CI) construction, is based on the following inferential rule. 
\begin{align}
&\textstyle \bar{\beta}=m^{-1}\sum_{l=1}^m\hat{\beta}^{(l)},\;  T=m^{-1}B+W\label{eqn:T}\\
&(\beta-\bar{\beta})T^{-1/2}\sim t_{\nu=(m-1)(1+mW/B)^2}, \label{eqn:inference}\mbox{ where}\\
&\mbox{$B=\sum_{l=1}^m(\hat{\beta}^{(l)}-\bar{\beta})^2/(m-1)$ (between-set variability)}\notag\\
&\mbox{$W=m^{-1}\sum_{l=1}^m w^{(l)}$ (within-set variability).}\notag
\end{align}\vspace{-12pt}

\section{\large{Privacy}-preserving Case \large{Surveillance} Data Release} \label{sec:surveillance}\vspace{-3pt}
WeWe present privacy-preserving release of three pandemic data types: subgroup case surveillance data (Section \ref{sec:surveillance}), case location data (Section \ref{sec:location}), and  CTNs (Section\ref{sec:ct}). In each case, we describe data characteristics, introduce methods for sanitization, conduct a simulation study to examine the impact of sanitization on statistical inference, and apply the method to a real data set when one is available. 

Case surveillance data are listing of cases, together with attributes associated with the cases, such as demographics,  exposure histories, etc.   Surveillance data  are crucial during the pandemic for monitoring and forecasting the spread of the disease, understand how COVID impacts the capacity of  healthcare systems, and provide necessary information to health authorities for quick decision making. Case numbers reported at different geographical scales by demographic groups such as age, gender, race and ethnicity provide valuable information for identifying risk factors and groups vulnerable to the disease and understanding the heterogeneity of the susceptibility to the disease.  On the other hand, publishing such granular information may lead to re-identification and disclosure risk, especially when data are sparse. This section focuses on publishing  granular case numbers with privacy guarantees. 

An example of case surveillance data is the COVID-19 death count data released by the U.S. CDC website. Table \ref{tab:death} show such a dataset we downloaded on May 24, 2022 (Table 2 at \url{https://www.cdc.gov/nchs/nvss/vsrr/covid19/health_disparities.htm}) with some minor modifications (we removed the race group ``unknown' and collapsed age groups $(0,4]$ and $[5,17]$ to a single $\!<\!18$ group, and age groups $[75,84]$ and $\!\ge\!85$ to a single $\!>\!74$ group). Table \ref{tab:death} contains two attributes -- age group and race/ethnicity; each  has 7 levels, leading to a $7\times7$ contingency table .  The sample size $n\!=\!998,262$,  assumed to be public information. 
\begin{table}[!htb]
\centering \vspace{-6pt}
\caption{\small{U.S. COVID-19} death counts by age and race/ethnicity (May 24, 2022) }\label{tab:death}
\vspace{-5pt}
\resizebox{1\textwidth}{!}{
\begin{tabular}{@{}l@{}ccccccc@{\hspace{2pt}}c@{}} 
\hline
Age  (ys) & \multicolumn{7}{c}{Race/Ethnicity}\\
\cline{2-8}
group  & NH White &NH Black &NH AIAN &NH Asian &NH NHPI &NH Mix &Hispanic & Total\\
\hline
<17 &387 &274 &15 &36 &11 &30 &303 &1056 \\
18-29 &2263 &1492 &187 &190 &49 &73 &2015 &6269 \\
30-39 &6661 &4144 &560 &558 &151 &157 &5919 &18150 \\
40-49 &17269 &8937 &1021 &1206 &265 &309 &13981 &42988 \\
50-64 &97418 &35753 &3198 &5312 &715 &952 &43657 &187005 \\
65-74 &141409 &37765 &2901 &7423 &501 &913 &38422 &229334 \\
>75 &380630 &54576 &3210 &16504 &449 &1380 &56711 &513460 \\
Total &646037 &142941 &11092 &31229 &2141 &3814 &161008 &998262 \\
\hline
\end{tabular}}
\resizebox{1\textwidth}{!}{
\begin{tabular}{l}
Race/ethnicity = `unknown' is not included in the table.\\
NH = Non-Hispanic; AIAN = American Indian or Alaska Native; NHPI = Native Hawaiian or Other Pacific\\  
 Islander;  ``Mix" means ``more than one race"\\
\hline
\end{tabular}}\vspace{-12pt}
\end{table}

 \subsection{Method}\label{sec:method1}\vspace{-3pt}
Publishing a privacy-preserving case number dataset can be formulated as releasing a multi-dimensional histogram or contingency table. The most straightforward approach for achieving DP when releasing a histogram and contingency table is the flat Laplace sanitizer, which injects  noise from the Laplace mechanism directly into each cell count in a histogram or contingency table; methods that achieve better utility in sanitizing count data for certain analyses exist, at the cost of more complicated implementation, such as  \citet{xiao2012dpcube, xu2013differentially, zhang2014towards,  li2018privacy, xiao2011ireduct,geng2015optimal, Eugenio, bowen2021differentially, hay2009boosting}, just to name a few.  Given that there exist many methods for sanitizing count data, many aiming at improving the utility of a certain type of analysis and not straightforward to implement, and our main goal is to demonstrate the application of DP in releasing count data in general without a specific downstream analysis task in mind, we employ the flat Laplace mechanism (we examined a couple of other approaches, but their performance is not as good as Laplace sanitizer in the in utility analysis. More details are provided in Section \ref{sec:summary1}).

In our problem setting, the Laplace sanitizer employs the Laplace mechanism in Definition \ref{def:Laplace}  to sanitize each cell count of the multidimensional histogram/ contingency table to be released. The $l_1$ global sensitivity of releasing a histogram/table is 1 (WLOS, we use the unbounded DP unless mentioned otherwise; the sensitivity is 2 if the bounded DP is used). Sanitized count in cell $k$ is $\tilde{y}_k \sim$ Laplace$(y_k,\epsilon^{-1})$  for $k=1,\ldots,K$ cells. 
Sanitized counts may  be negative as the support of the Laplace distribution is the real line. There are two ways to deal with this problem -- to replace negative values by 0, and to re-draw until the sanitized value is non-negative \citep{liu2019bounding}. In either case, normalization would be needed if the total sample size $n$ is fixed. Real non-negative sanitized counts can be rounded to obtain integer counts without compromising privacy due to the immunity to post-processing property. 

To obtain sanitized counts for a lower-dimensional histogram/contingency table from the sanitized histogram/table at a more granular level, one may sum sanitized counts over corresponding cells to obtain cells counts in the lower-dimensional histogram/table. Per the immunity to post processing  property, the summed counts are also  privacy-preserving, but are subject to a larger sanitization variability  since each contains the sum of multiple independent noise terms.  

\subsection{Simulation Study} 
We use a simulation study to study how DP sanitization affects statistical inference based on sanitized count data.  We simulated 1,000 datasets from $y_k\sim$ multinomial($n, p_k$), where $p_k=\lambda_k/(1+\lambda_k)$, $\log(\lambda_k)=\beta_0+\beta_1 x_{k1}+\beta_2 x_{k2} + \beta_3 x_{k3}+ \beta_4 x_{k1}x_{k2} + \beta_5 x_{k1}x_{k3} + \beta_6 x_{k2}x_{k3}$ for $k=1,\ldots,8$ and $X_1=\{0,1\}, X_2=\{0,1\}, X_3=\{0,1\}$ are binary attributes. In each dataset, we sanitize $\mathbf{y}=\{y\}_{k=1,\ldots,8}$ via the flat Laplace sanitizer independently for $m=3$ times to obtain differentially private $\tilde{\mathbf{y}}^{(l)}$ and $l=1,\ldots,m$, each at a privacy budget of $\epsilon/m$, where $\epsilon$ is the total privacy budget. We examine two  sample sizes at $n=200$ and $n=1,000$ and four privacy loss parameters at $\epsilon=0.5,1,2$ and 5.  We assume the total sample size $n$ is fixed and normalize the raw sanitized counts from  the flat sanitizer via $n\tilde{y}_k^{(l)}/\sum_l\tilde{y}_k^{(l)}$. For utility check, we run the loglinear model $\log(\lambda_k)=\beta_0+\beta_1 x_{k1}+\beta_2 x_{k2} + \beta_3 x_{k3}+ \beta_4 x_{k1}x_{k2} + \beta_5 x_{k1}x_{k3} + \beta_6 x_{k2}x_{k3}$ for $k=1,\ldots,8$,  assuming $\tilde{\mathbf{y}}^{(l)}_k\!\sim$ Poisson($\lambda_k$), on each set of sanitized data to obtain inference on $\beta_1,\ldots,\beta_6$ using Eqs \eqref{eqn:T} and \eqref{eqn:inference}.  For comparison, we also run the same loglinear model on the original $\mathbf{y}$. 

\begin{figure}[!htb]
\vspace{-9pt}\centering
\includegraphics[width=1\textwidth]{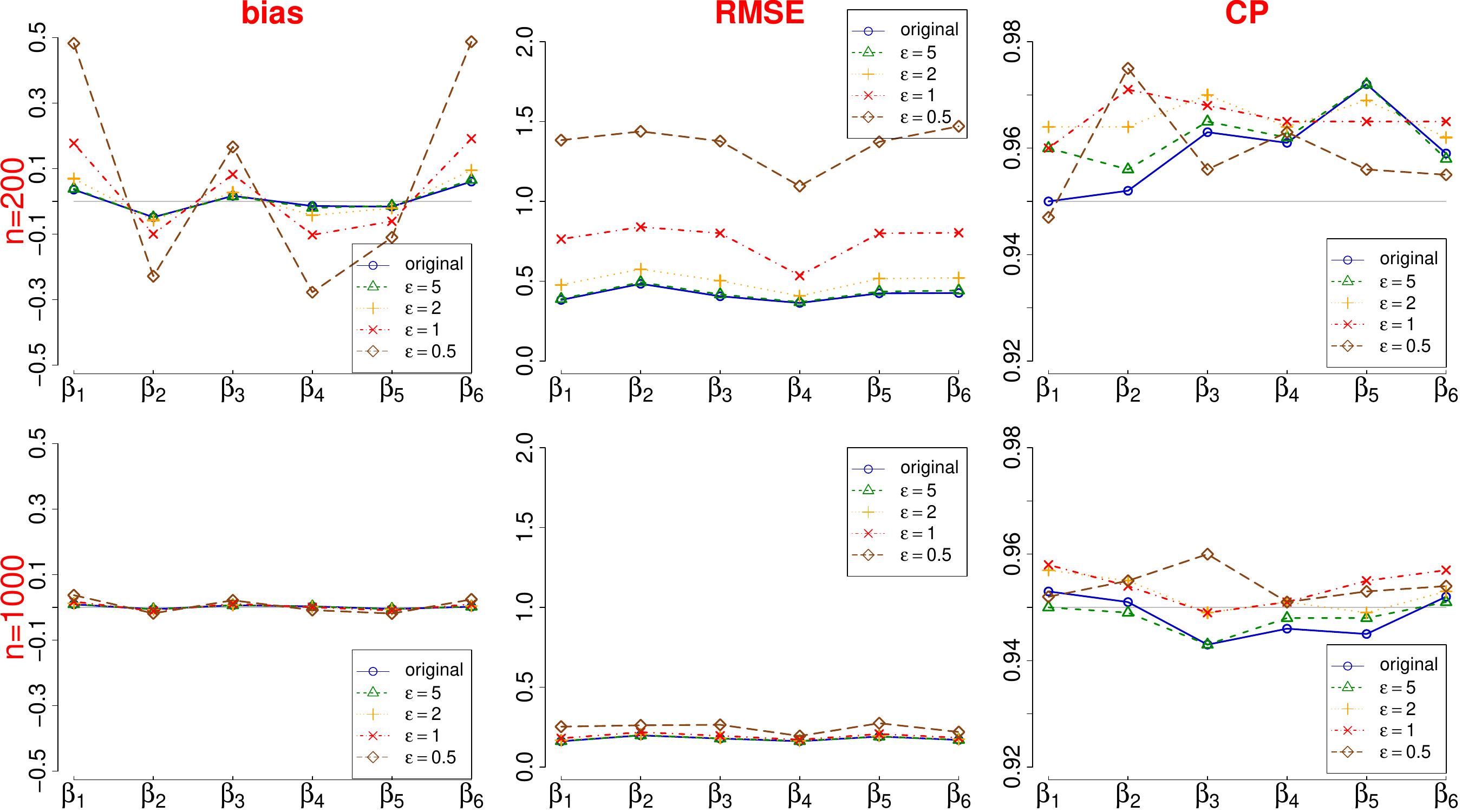}
\vspace{-12pt} \caption{Privacy-preserving inference in log-linear model on sanitized counts obtained via the flat Laplace sanitizer in simulated data (1000 repeats)} \label{fig:sim1} \vspace{-12pt}
\end{figure}

The results are presented in Figure \ref{fig:sim1} and the main observations are summarized as follows. The smaller $\epsilon$ or $n$ is, the more impact the DP procedure has on the inference; i.e., larger bias and larger root mean squared error (RMSE).  Regardless of $n$ or $\epsilon$,  the coverage probability (CP) of the 95\% CIs is always at the nominal level. At $n=1,000$, the inference is barely affected by the DP sanitization even for $\epsilon=0.5$. At $n=200$, the bias is noticeable with relatively large RMSE for $\epsilon=0.5$, acceptable at $\epsilon=1$, and almost ignoble for  $\epsilon>1$, compared to the original inference. 

\vspace{-3pt}\subsection{Application to CDC case surveillance data}

We apply the flat Laplace sanitized to the CDC in Table \ref{tab:death}.  If released data are not used for statistical inference or uncertainty quantification, we may release a single sanitized tabular dataset ($m=1$). Let $\tilde{y}_k=y_k+e_k$,  where $e_k\sim$ Laplace($0,\epsilon^{-1}$),  for $k=1,\ldots,49$ independently. Since $n=998,262$ is public knowledge, the sanitized $\tilde{y}_k$ is normalized as in $\tilde{y}_k\leftarrow n\tilde{y}_k/\sum_k{\tilde{y}_k}$ to keep the total $n$ at $998,262$.  An example sanitized dataset  at $\epsilon=0.5$ is given in Table \ref{tab:death.1}.  There is some fluctuation in each cell count due to the sanitization, as expected. The column and row marginals are calculated by summing over the corresponding cell counts after santization.
\begin{table}[!htb]
\centering \vspace{-4pt}
\caption{Flat Laplace sanitized ($\epsilon=0.5, m=1$) US COVID-19 death counts by age group and race/ethnicity on May 24, 2022}\label{tab:death.1}
\vspace{-6pt}
\resizebox{1\textwidth}{!}{
\begin{tabular}{lcccccccc} 
\hline
Age  (ys) & \multicolumn{7}{c}{Race/Ethnicity}\\
\cline{2-8}
group  & NH White &NH Black &NH AIAN &NH Asian &NH NHPI &NH Mix &Hispanic & Total\\
\hline
<17 &385 &271 &14 &37 &8 &29 &308 &1052 \\
18-29 &2258 &1491 &186 &198 &49 &72 &2009 &6263 \\
30-39 &6664 &4140 &562 &558 &145 &156 &5928 &18153 \\
40-49 &17269 &8937 &1021 &1202 &266 &299 &13982 &42976 \\
50-64 &97421 &35753 &3195 &5311 &713 &952 &43658 &187003 \\
65-74 &141413 &37766 &2897 &7427 &501 &914 &38425 &229343 \\
>75 &380642 &54577 &3209 &16505 &449 &1379 &56712 &513472 \\
Total &646053 &142935 &11084 &31238 &2130 &3801 &161021 &998262 \\
\hline
\end{tabular}}
\resizebox{1\textwidth}{!}{
\begin{tabular}{l}
Race/ethnicity = 'unknown' is not included in the table.\\
NH = Non-Hispanic; AIAN = American Indian or Alaska Native; NHPI = Native Hawaiian or Other Pacific \\  
Islander;  "Mix" means "more than one race"\\
\hline
\end{tabular}}\vspace{-11pt}
\end{table}

\begin{figure}[!htb]
\vspace{-3pt}\centering
\includegraphics[width=1\textwidth, trim= 0cm 0cm 0cm 26.6cm, clip]{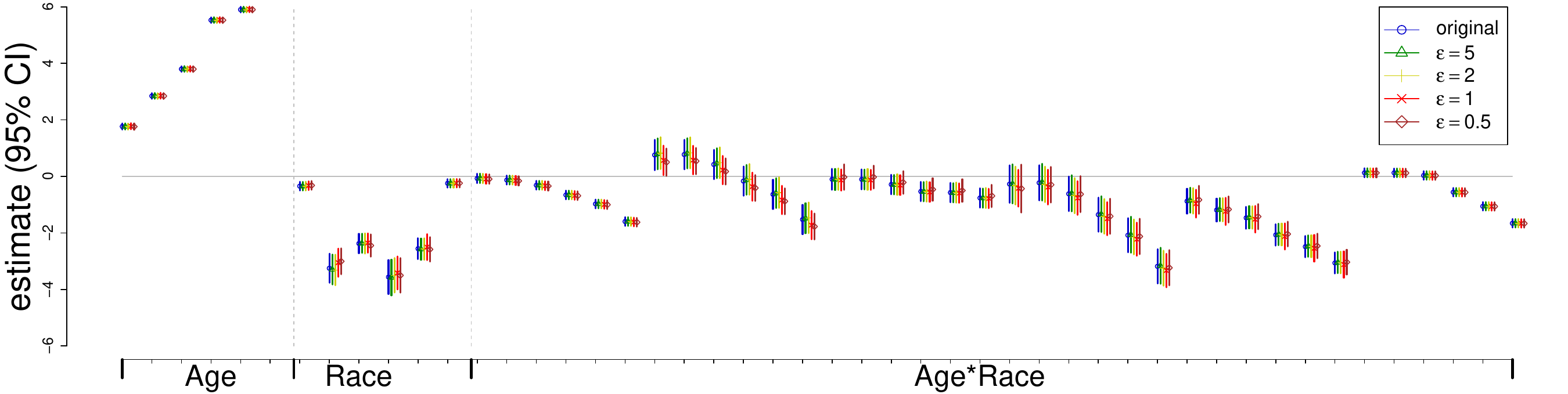}\vspace{-3pt} 
\caption{Privacy-preserving results from the log-linear model on sanitized CDC COVID-19 death data via the flat Laplace sanitizer}\label{fig:cdc} \vspace{-12pt}
\end{figure}

If the data will be used for statistical inference, we can use the MS approach to release multiple sets of sanitized tables. 
We set $m=3$  and sanitized $y_k$ with noise from Laplace($0,\epsilon/m)$  independently  to obtain 3 sets of sanitized $\tilde{y}_k^{(l)}$ for $l=1,2,3$. Some example sanitized data are provided in the supplementary materials. For the statistical analysis on the sanitized data, we fitted a 2-way loglinear model with covariates age group and race/ethnicity  (other analysis can also be run, such as logistic regression, Chi-squared test). There are  48 regression coefficients -- 6 associated with age ($<18$ years is the reference group), 6 associated with race (non-Hispanic white is the reference group), and 36 parameters representing the interaction between the two.  The estimates of the regression coefficients are presented in Figure \ref{fig:cdc}.  In summary, the privacy-preserving inferences based on the sanitized counts are similar to the original inference at all $\epsilon$ values, largely due to the large sample size of the data. 

\subsection{Summary}\label{sec:summary1}\vspace{-3pt}
Case number data with granular information permits more complicated analysis and helps us understand better the pandemic, such as quantifying the effects of risk factors for COVID-19 as demonstrated in Figure \ref{fig:cdc}). We demonstrate via a simulation study and a real data application that useful privacy-preserving can be achieved, especially when $n$ is large or people are willing to sacrifice some privacy ($\epsilon$ is not too small).  The results also suggest the flat Laplace sanitizer can be an effective approach for that purpose, despite its simplicity.  

Though we focus on the flat Laplace sanitizer for demonstration purposes, we also run a couple of other methods that sanitize count data in a hierarchical manner in the simulation study and the case study. The two approaches are -- - the  universal histogram (UH) approach \citep{hay2009boosting} and its extension UH-proportion or simply UHp  that we extend UH for the case where the total sample size of the released data is fixed and public.  The descriptions of the UH and UHp approaches,  the details of their implementation, and the results from the simulation study and the case study are presented in the supplementary materials.
In summary,  UHp delivers comparable performance to the flat sanitizer in bias and RMSE for most of the parameters in the simulation study, but has slight under-coverage  at $\epsilon=$ 1 and 0.5.  UH performs the worst (largest bias, RMSE, and some notable under-coverage).  In the case study, there is some discrepancy between the privacy-preserving point estimates vs the original for both UH and UHp. For UH, some  CIs are noticeably wider than the original, mostly in the race/ethnicity groups that are relatively small in size. 

\section{\large{Privacy}-preserving Release of \large{Case} Location Data} \label{sec:location}\vspace{-3pt}
When a person is diagnosed with COVID-19, health authorities may interview the person for his or her whereabouts and location history in the past few weeks \citep{locationhistory, publishlocationinformation}.  Patient's location data are critical for health authorities to  take measures to limit the spread of the disease. With individual-level location data, researchers can conduct spatial data analysis such as  using  point process models  to understand the spatial trend of the cases or generating  COVID-19 hot spot heat maps.
However, location information, if shared as is, may cause serious privacy risk for the patients and can even lead to cyber-bullying \citep{locationprivacy}.

We examine a privacy-preserving approach to releasing location data  based on  GI.  We focus on releasing cross-sectional location data at a given time point rather than travel trajectories \citet{liu2021privacy}, which is a topic for future research.  Even though released data is cross-sectional, they can be released on a regular time basis, e.g., every day or every 3 days, allowing temporal examination of certain trends.

\begin{figure}[!htb]\vspace{-9pt}\centering
\includegraphics[width=0.3\textwidth]{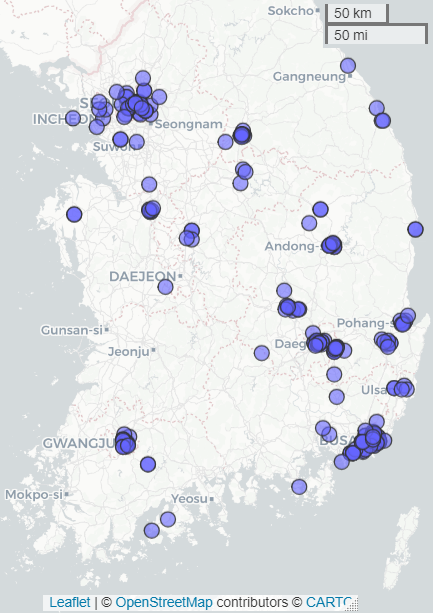}
\includegraphics[width=0.5\textwidth, trim={3cm 3cm 1cm 3cm},clip]{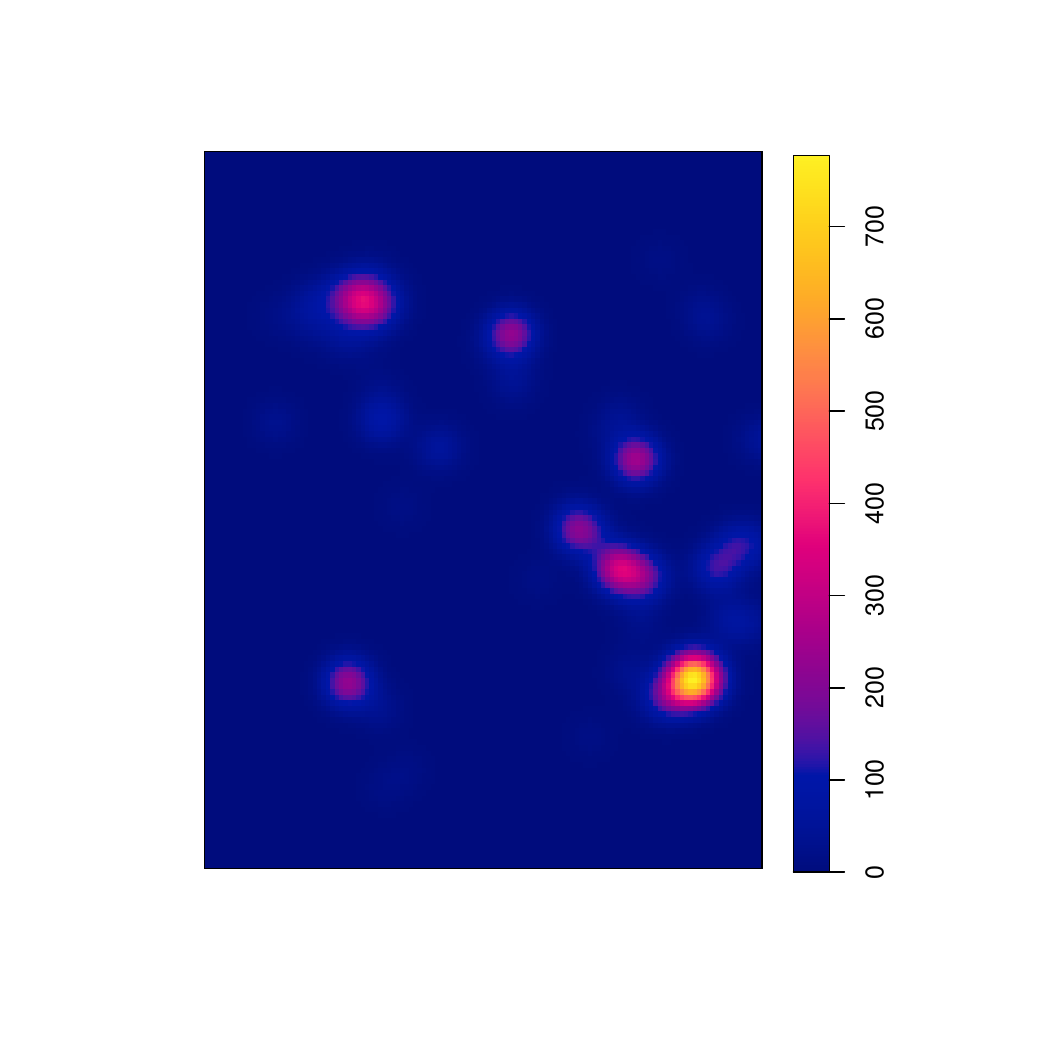}\\
\hspace{-0.5in}(a) actual location \hspace{0.6in} (b) hot spot heat map  \vspace{-6pt}
\caption{Locations of 121 COVID-19 patients on Feb 20, 2020 in South Korea}\label{fig:SKmap}\vspace{-12pt}
\end{figure}

An example of location data  is given in Figure \ref{fig:SKmap}, which shows the locations of 121 COVID-19 patients on Feb 20, 2020 in South Korea. The data can be found in file ``patientroute.csv'' at \url{https://www.heywhale.com/mw/dataset/5e797e9e98d4a8002d2c92d3/file}. The number of locations per subject ranges from 1 to 11; about 50\% (62 out of 121) has one location, 34.7\% has  2 or 3 locations, and the rest 14\%  have $\ge4$  locations (one person has 11 locations; all within the city of Gwangju). The  timestamp  information in hours, minutes, and dates is not available in the dataset.  

\subsection{Method}\vspace{-3pt}
The approach we propose for releasing privacy-preserving location information is \textit{the \D} \citet{liu2021privacy}, based on the GI concept. The main idea behind \D, as suggested by the name, is to release $m\ge1$ sanitized versions of the true location $P$ via the planar Laplace mechanism so to satisfy GI guarantees. The privacy budget per location $\epsilon$ is split into $m$ portion for $m\ge2$, $\epsilon/m$ per release.  Similar to case surveillance data, the main reason for releasing multiple perturbed locations ($m\ge2$) is to provide a way to quantify sanitization uncertainty and draw statistical inferences using the MS approach.  

To generate a sanitized location $(x^*,y^*)$ given the original location coordinates $(x,y)$, we apply the planar Laplace mechanism in Eq \eqref{Eq:xsys}, with $\epsilon$ replaced by $\epsilon/m$. $\epsilon$ is the per-unit-distance privacy loss, where the unit distance is supplied by the data curator and can be any value deemed appropriate for the task at hand, such as 1 meter, 10 meters, 0.5 miles, etc (generally speaking, the choice  depends on location type, area, among other considerations). 
\subsection{Simulation Study}
To evaluate  statistical utility of sanitized locations via \D, we conduct a simulation study. We simulated 1,000 sets of location data in a square area of $[0, 1]\times[0, 1]$ from an inhomogeneous Mat\'{e}rn cluster point process with the radius of the clusters  at $0.03$ and the non-stationary  log-density 
$\log(\lambda(x,y;\bs\beta))= \beta_0+\beta_1x+\beta_2y+\beta_3x^2+\beta_4y^2+\beta_5xy$, where $x$ and $y$ are coordinates and $\bs\beta=(\beta_0,\ldots,\beta_5)=(4.53, 3.30, 3.43, -0.27, 1.58, 2.24)$.  The number of locations ranges from 769 to 1217 across the 1,000 repeats with an average of 970. In each simulated dataset, we sanitized each location with the planar Laplace mechanism in Eq \eqref{Eq:xsys} at $\epsilon=5, 2, 1, 0.5$ per 0.01 unit 
and $m=3$. We assume $[0, 1]\times[0, 1]$ is public information and sanitized locations thus should fall within $[0, 1]\times[0, 1]$. On the other hand, the planar Laplace mechanism can generate an infinite $r$ and any angle $\in[0,2\pi]$. To honor the location boundaries, we set sanitized $x^*<0$ at $0$ and at 1 if it is $>1$; similarly for sanitized $y^*$. We then fitted the inhomogeneous Mat\'{e}rn cluster point process model above and applied the inferential  rule in Eq \eqref{eqn:inference} to obtain inference on $\bs\beta$. The data simulation and analysis were conducted using R package \texttt{spatstat.core} \citep{baddeley2005spatstat}.

The results are presented in Table \ref{tab:location}. In summary, the inferences at $\epsilon=5$ and $\epsilon=2$ are comparable to the original -- close-to-0 bias, similar RMSE as the original, nominal converge at $\epsilon=5$ and slight under-coverage at $\epsilon=2$. At $\epsilon=1$ and $\epsilon=0.5$, the bias is notable; the RMSE values are similar to the original at $\epsilon=1$, but much larger at $\epsilon=0.5$; the CP is around 83\% to 85\% at $\epsilon=1$ and ranges from 60\% to 88\% at $\epsilon=0.5$. The moderate to severe under-coverage is largely due to the bias in the $\bs\beta$ estimates, which in turn may be attributed to the bounding applied to the sanitized locations.  Bounding sanitized values can lead to biased inference \citep{liu2019bounding}.
\begin{table}[!htb]
\centering \vspace{-9pt}
\caption{Privacy-preserving inferences of Mat\'{e}rn cluster point process model on simulated location data (1,000 repeats, $m=3$) } \label{tab:location}\vspace{-3pt}
\resizebox{0.68\linewidth}{!}{
\begin{tabular}{llccccc}
\hline
metric & parameter  & original & $\epsilon=5$ & $\epsilon=2$ & $\epsilon=1$ & $\epsilon=0.5$ \\
\hline
 & $\beta_0$ &-0.029 &-0.022 &0.016 &0.142 &0.571 \\
 & $\beta_1$ & 0.065 &0.052 &-0.022 &-0.279 &-1.180 \\
bias & $\beta_2$ &0.031 &0.014 &-0.074 &-0.374 &-1.389 \\
& $\beta_3$ &-0.085 &-0.077 &-0.028 &0.154 &0.801 \\
& $\beta_4$ &0.034 &0.038 &0.060 &0.124 &0.337 \\
& $\beta_5$ &-0.037 &-0.024 &0.048 &0.303 &1.160 \\
\hline
& $\beta_0$ &0.466 &0.465 &0.459 &0.457 &0.680 \\
& $\beta_1$ &1.234 &1.232 &1.211 &1.189 &1.549 \\
RMSE & $\beta_2$ &1.164 &1.162 &1.152 &1.166 &1.693 \\
& $\beta_3$  &1.006 &1.003 &0.986 &0.958 &1.159 \\
& $\beta_4$ &0.944 &0.943 &0.934 &0.898 &0.838 \\
& $\beta_5$  &0.985 &0.982 &0.972 &0.989 &1.431 \\
\hline
& $\beta_0$ &0.948 &0.940 &0.925 &0.841 &0.599 \\
& $\beta_1$&0.938 &0.932 &0.914 &0.845 &0.719 \\
CP& $\beta_2$ &0.957 &0.952 &0.935 &0.851 &0.640 \\
& $\beta_3$&0.938 &0.929 &0.909 &0.842 &0.769 \\
& $\beta_4$&0.941 &0.934 &0.908 &0.840 &0.878 \\
& $\beta_5$&0.947 &0.939 &0.916 &0.827 &0.638 \\
\hline
\end{tabular}}\vspace{-9pt}
\end{table}

\subsection{Application to South Korea case location data}\vspace{-3pt}
We apply the \D\hspace{1pt} to the real South Korean case location dataset (Figure \ref{fig:SKmap}(a)) to release privacy-preserving locations at $\epsilon=5, 2, 1, 0.1$ per 2 miles per individual. For an individual who has more than one location record, we further divided $\epsilon$ by the number of locations for that individual. That is, if an individual has $h$ original location data points  and we release $m$ sanitized locations for each location at a privacy budget of $\epsilon/(mh)$.  Similar to the simulation study, we honor the fact that all cases are in South Korea and bounded sanitized locations within a rectangular that approximates the shape of South Korea, in a similar fashion as done in the simulation study. 

\begin{figure}[!htb]\centering\vspace{-6pt}
$\epsilon=5$\hspace{0.7in}$\epsilon=2$\hspace{0.7in}$\epsilon=1$\hspace{0.75in}$\epsilon=0.5$
\small $m\!=\!1$\raisebox{-0.75in}{
\includegraphics[width=0.225\textwidth, trim={3.5cm 3cm 2.5cm 3cm},clip]{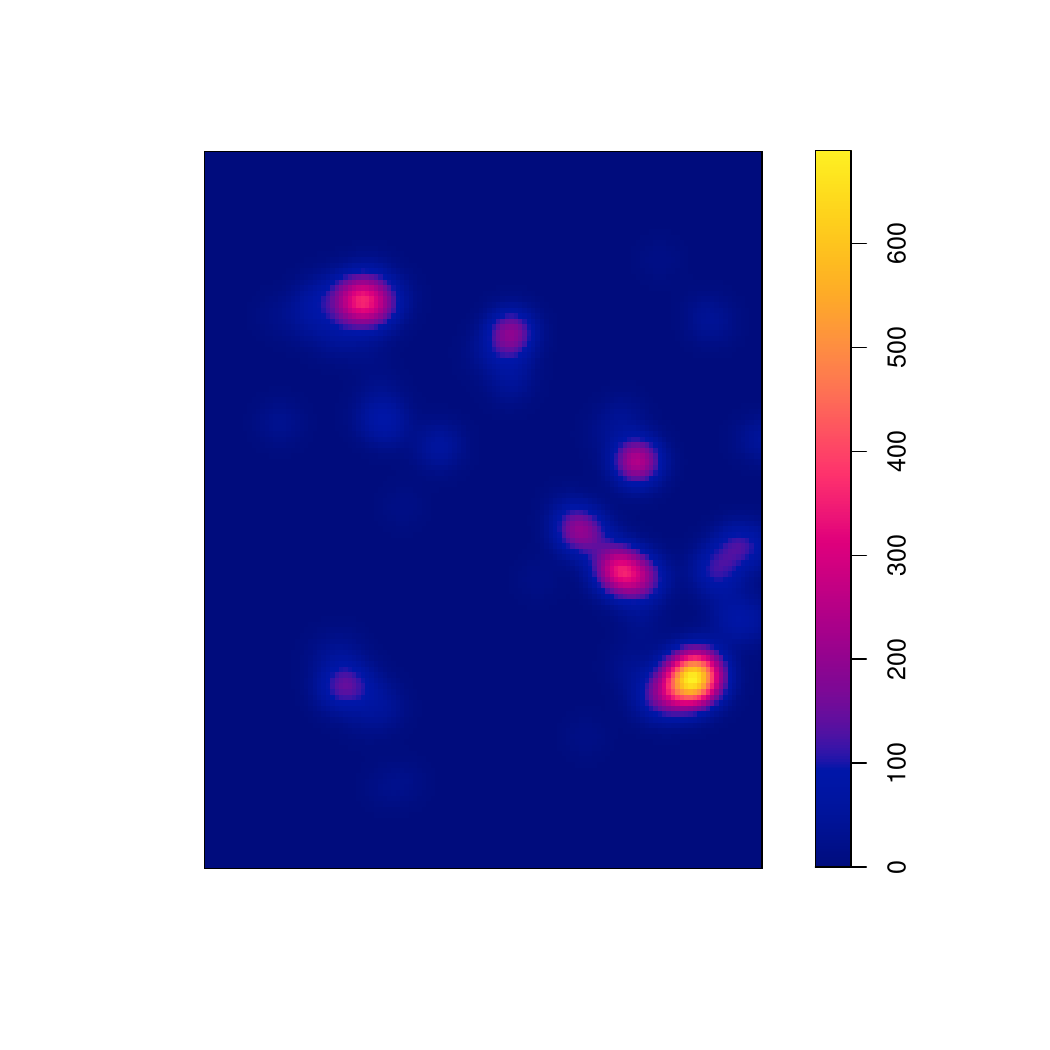}
\includegraphics[width=0.225\textwidth, trim={3.5cm 3cm 2.5cm 3cm},clip]{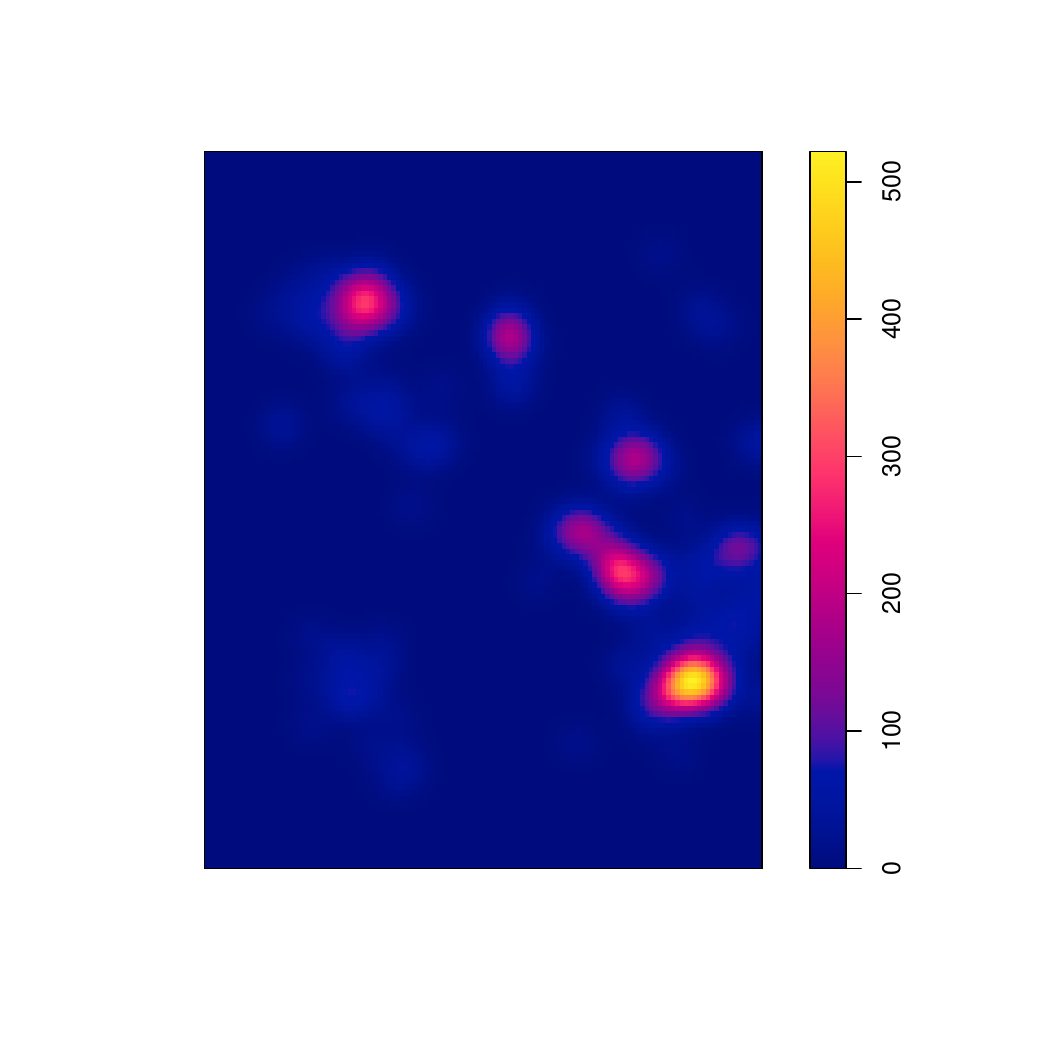}
\includegraphics[width=0.225\textwidth, trim={3.5cm 3cm 2.5cm 3cm},clip]{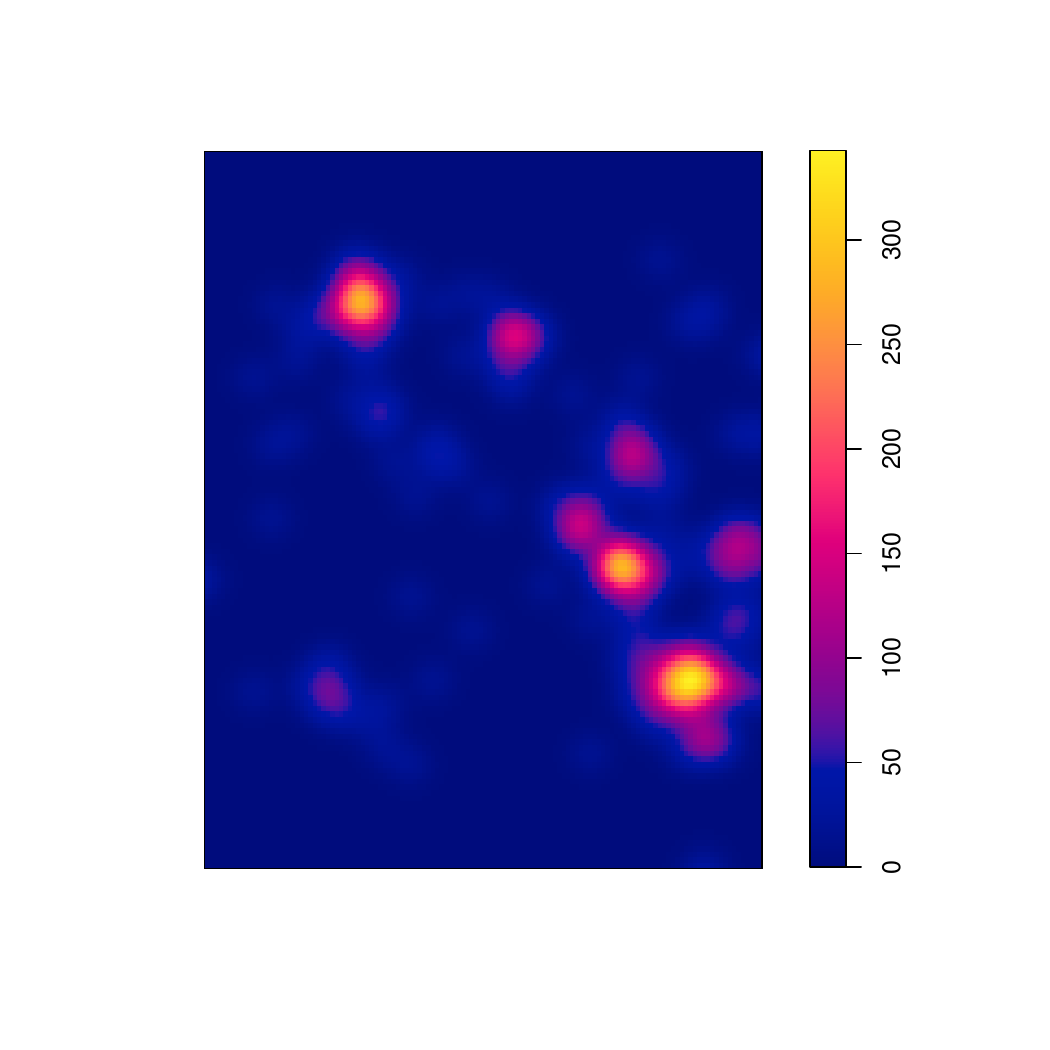}
\includegraphics[width=0.225\textwidth, trim={3.5cm 3cm 2.5cm 3cm},clip]{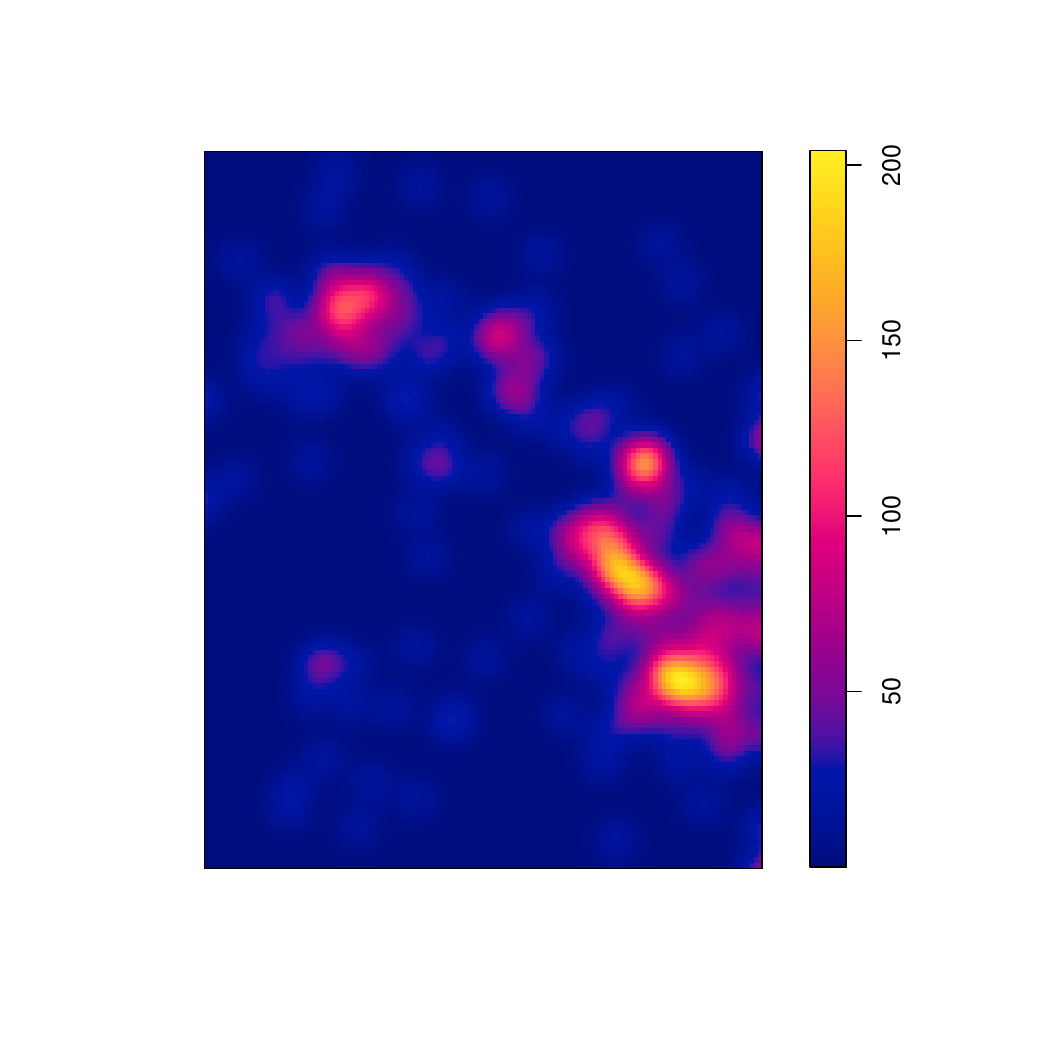}}\\
\vspace{3pt}
\small $m\!=\!3$\raisebox{-0.75in}{
\includegraphics[width=0.225\textwidth, trim={3.5cm 3cm 2.5cm 3cm},clip]{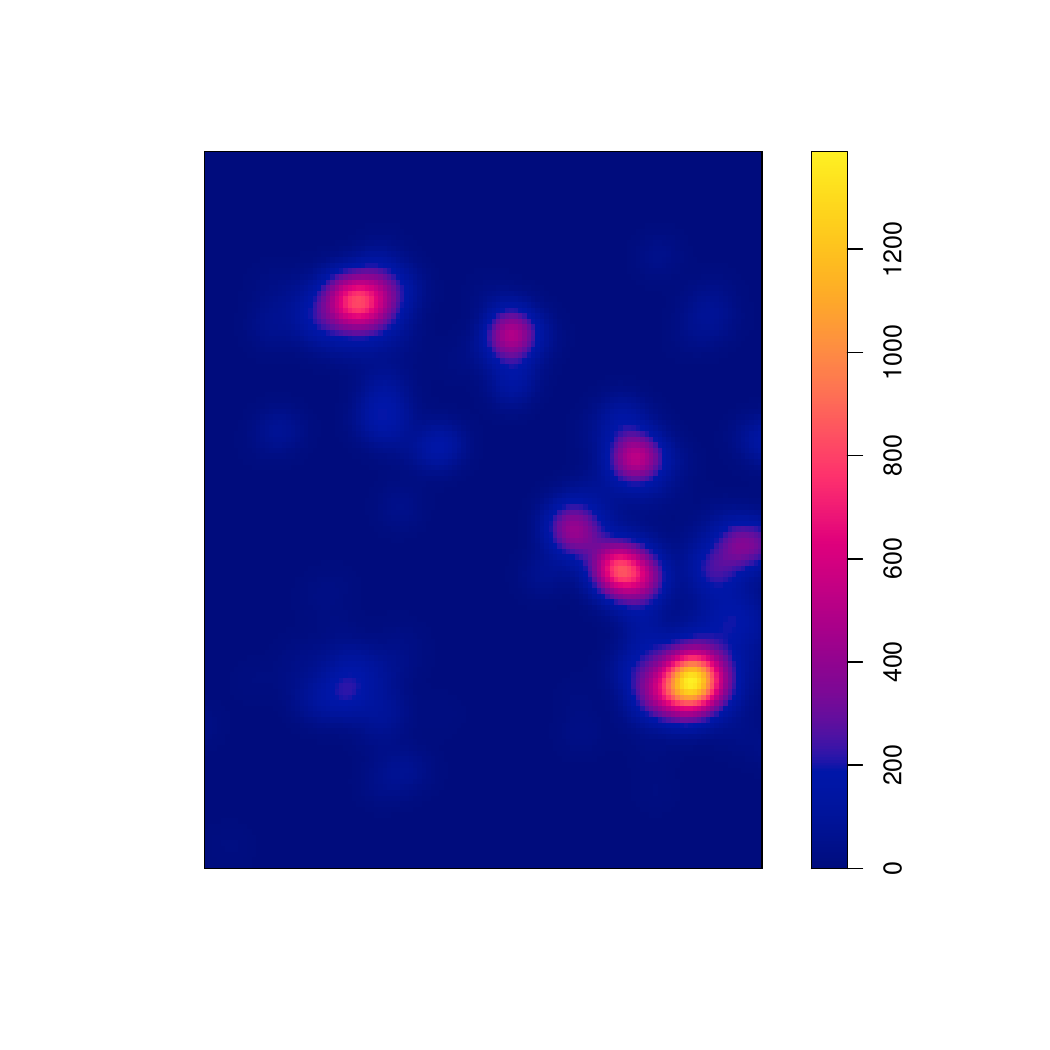}
\includegraphics[width=0.225\textwidth, trim={3.5cm 3cm 2.5cm 3cm},clip]{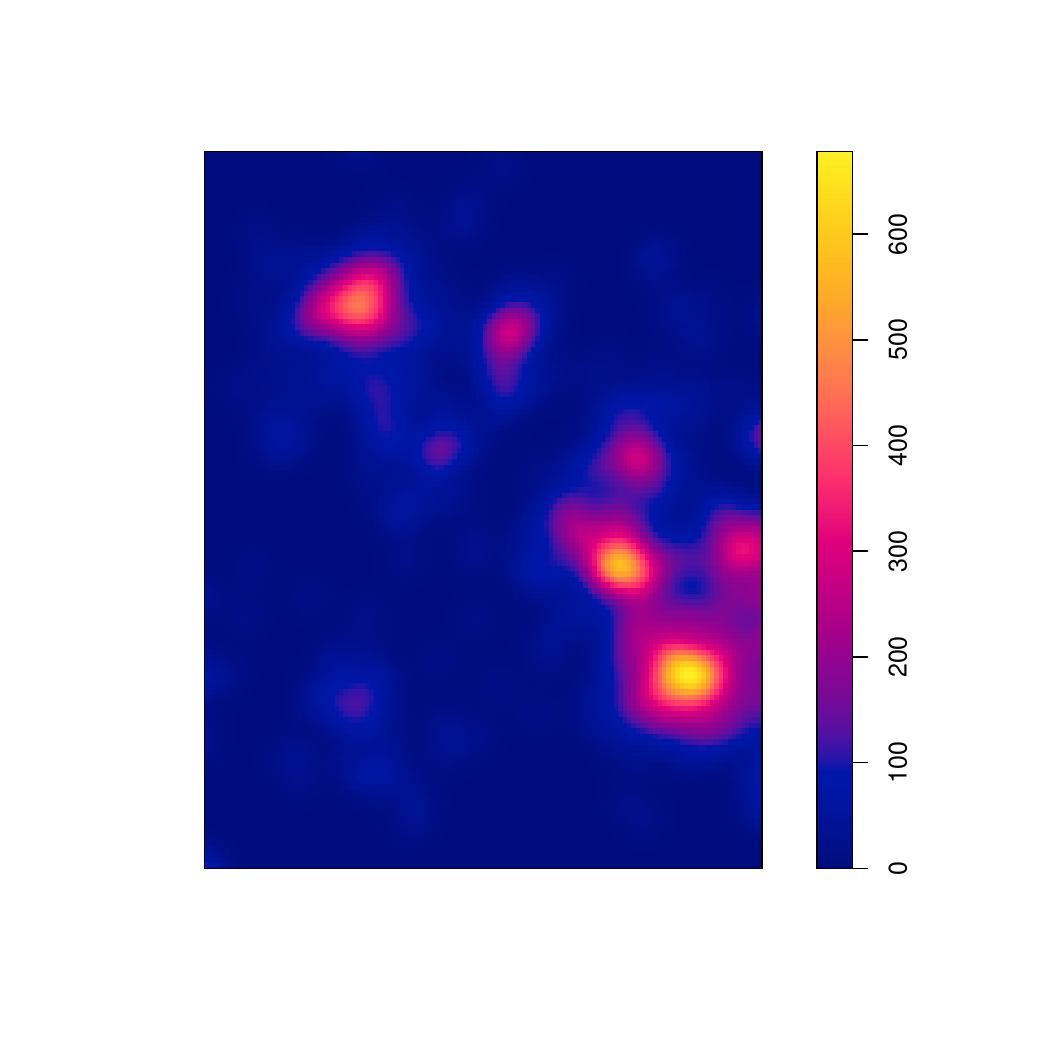}
\includegraphics[width=0.225\textwidth, trim={3.5cm 3cm 2.5cm 3cm},clip]{figure R1/SKheatmap256m3_2_d0_0.025.pdf}
\includegraphics[width=0.225\textwidth, trim={3.5cm 3cm 2.5cm 3cm},clip]{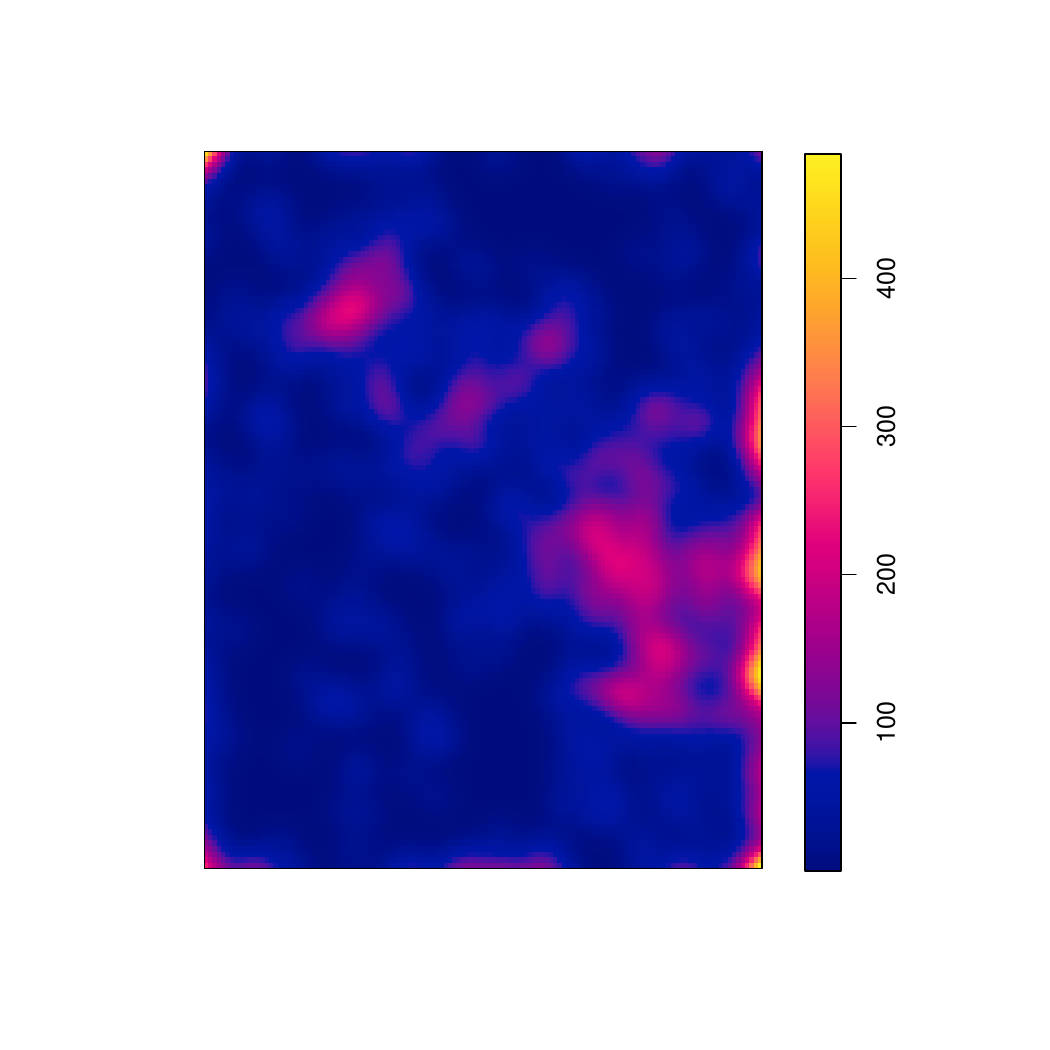}}\vspace{-3pt}
\caption{Privacy-preserving COVID-19 hot spot heat maps  in South Korea on Feb 20, 2020} \label{fig:heatmap} \vspace{-12pt}
\end{figure}

We used two analyses to check the utility of the sanitized locations: to generate hot spot heat maps and to fit a point process model. We set $m=3$ in both analyses but also examined $m=1$ in the former as it does not involve statistical inference.  The privacy-preserving heat maps are displayed in Figure \ref{fig:heatmap} with the same smoothing bandwidth as in Figure \ref{fig:SKmap}(b). 
The privacy-preserving hot spot heat maps are very similar to the original heat map in Figure \ref{fig:SKmap}(b) at $\epsilon\ge1$ for both $m=1$ and $m=3$ and are a bit noisy at  $\epsilon=0.5$ especially when $m=3$; but the major hot spots (the cities of Busan, Seoul, and Daegu) are  preserved at $\epsilon=0.5$ for $m=1$. In summary, for the purposes of generating heat maps, $m=1$ is sufficient and each sanitized location is less noisy compared to  using $m=3$ especially at small $\epsilon$.

We fitted an inhomogeneous Mat\'{e}rn cluster point process model with log-density $\log(\lambda(x,y;\bs\beta))=\beta_0+\beta_1x+\beta_2y$ on the original data and the sanitized data. For this analysis, we randomly selected one  location if an individual has multiple original location records, resulting in one original location per individual. We applied the inferential rule in Eqs \eqref{eqn:T} and \eqref{eqn:inference} to obtain the point estimates and 95\% CIs for $(\beta_0,\beta_1,\beta_2)$.  The results are presented in Table \ref{tab:SKinference}. In general, the privacy-preserving inferences are similar to the original, especially for $\beta_1$ and $\beta_2$ that quantify the linear trends of COVID intensity along the $x$ and $y$ coordinates, respectively. In addition, the privacy-preserving point estimates are robust to $\epsilon\ge1$ and some notable deviation from the original is only seen at $\epsilon=0.5$. A surprising observation is the shrinkage in the CIs as $\epsilon$ decreases for $\epsilon<5$, implying the inferences become more precise, at least for the range of the examined $\epsilon$ values, though the statistical insignificance remains unchanged across $\epsilon$. The shrinkage is counter-intuitive as one would expect the inferences get less precise as the locations are  perturbed more at smaller $\epsilon$.  As $\epsilon$ decreases, the sanitized locations are more scattered (Figure \ref{fig:heatmap}) and the likelihood of  a sanitized location being bounded at the boundary also  increases, which may affect the within and between components of the total variance in Eq \eqref{eqn:T}.  More research is needed to understand precisely how the variability is affected by the sanitization and the bounding constraint. 
\begin{table}[!htb]\vspace{-6pt}
\caption{{\small Privacy-preserving Mat\'{e}rn} cluster point process model parameter estimates based on sanitized locations in the {\small South Korea}  location data ($m\!=\!3$).} \label{tab:SKinference} 
\centering\vspace{-6pt}
\resizebox{\textwidth}{!}{
\begin{tabular}{@{}c@{\hspace{4pt}}c@{\hspace{4pt}}c@{\hspace{4pt}}c@{\hspace{4pt}}c @{\hspace{4pt}}c@{}}
\hline
 & \multicolumn{5}{c}{estimate (95\% CI)} \\
\cline{2-6}
&  original & $\epsilon=5$ & $\epsilon=2$ & $\epsilon=1$ &$\epsilon=0.5$\\
\hline
$\beta_0$ & -64.2 (-153.5, 25.1) & -65.1 (-157.0, 26.9) & -63.0 (-147.0, 21.0) & -63.8 (-140.6, 13.0) & -57.5 (-129.8, 14.7) \\
$\beta_1$ &  0.51 (-0.17, 1.19) &  0.52 (-0.18, 1.21) & 0.50 (-0.14, 1.14)  & 0.50  (-0.08, 1.08)  & 0.44 (-0.10, 0.99)  \\
$\beta_2$ & 0.03 (-0.50, 0.56) & 0.03 (-0.52, 0.59) & 0.03 (-0.48, 0.54) & 0.05 (-0.42, 0.51) & 0.07 (-0.39, 0.53) \\
\hline
\end{tabular}}\vspace{-12pt}
\end{table}

\subsection{Summary}\vspace{-3pt}
The doppelganger method releases location data with privacy guarantees. The simulation study and the case study suggest the method can preserve important statistical signals in the original data at a relatively low-level cost of privacy. The method would be particularly useful for protecting location privacy when sharing information at a local level or releasing hot spot maps on a relatively fine scale. The finer the scale is, the more sparse the data become, the higher the privacy risk for re-identification from releasing location data, and the greater the need for effective  privacy protection approaches, but also the noisier released sanitized locations. As the scale gets coarser, say at the city,  regional, state, or national levels, the information released by the \D$\;$ can be very similar to the original location information.

\section{Privacy-preserving Sharing of Contact Tracing Networks (CTNs)} \label{sec:ct}
Contact tracing (CT) is an effective approach for curbing the spread of COVID-19 during the pandemic. CT can be carried out manually by human  tracers or digitally via GPS or Bluetooth devices. CTNs, constructed from CT data, can be regarded as a type of social networks with individuals being the nodes and an edge between two people representing a close contact between them (e.g., within 6 feet of each other for a cumulative total of 15 minutes or more over a 24-hour period).   CTNs are of research interest as they provide valuable information to better understand how physical proximity affects the spread of the disease and human contact behaviors during the pandemic, among others. However, sharing CTNs as is has privacy concerns as adversaries may link a CTN with other databases or use background knowledge to  infer who were infected with COVID-19 and tell who were close physically (appearing in the same place at the same time)  based on the edge information in a CTN.  

CT data are only collected as needed -- that is, when a person is diagnosed positive for COVID-19. Therefore, a CTN only contains COVID-positive individuals and their close contacts. That said,  CTNs can be constructed in different ways from CT data, and they can be complex and large as people are mobile and may show up in various places at different times.   We focus on CTNs constructed for a pre-defined population during a pre-specified period time  (e.g., employees  in an organization or students in a school in one day, 2 weeks, or 1 month, etc). For example, suppose the time period is one day, starting at noon on June 1 2020 ending at noon on the next day and the population is all students at a college. If a  COVID-positive student named Tom was in a dining hall from noon to 1pm on June 1, 2020 and had 2 close contacts, at the library from 1:30pm to 5pm and had 1 close contact, and in his dorm from 5pm to noon next day and had 5 close contacts, then Tom and all his 8 close contacts are included in the CTN, along with 8 edges, representing the 8 close contacts. 
We consider privacy-preserving release of CTNs with relational information only in this study; releasing CTNs with nodal attributes (such as demographic information or location information) with privacy guarantees is a topic for future research.

\subsection{Method}\label{method3}

We examine a few approaches for releasing privacy-preserving CTNs 
and present one approach, DP-ERGM, in the main text and include the other two in the supplementary materials. 
DP-ERGM stands for Differentially Private network synthesis via Exponential Random Graph Model \citep{DPERGM}.  The DP-ERGM procedure can be regarded as an application of the model-based differentially private synthesis (MODIPS) approach \citep{liu2016model} to graph data with ERGM as the synthesis model.  ERGMs are a family of popular statistical models for analyzing network data \citep{Snijders:2006, Robins:2007}. Denote by $\mathbf{e}$ the adjacency matrix in a  network ($e_{ij}=1$ if an edge exists between node $i$ and node $j$,  $e_{ij}=0$ otherwise). ERGMs model the conditional distribution of $\mathbf{e}$ as 
\begin{equation}\label{eq:ergm}
p(\mathbf{e} |\bs\theta) = \frac{\exp \big\{ \boldsymbol{\theta}^{T} \mathbf{S}(\mathbf{e}) \big\}}{K(\bs{\theta})} \mbox{ with }\textstyle  K(\bs{\theta})= \sum_{\mathbf{e}'} \exp \left\{ \bs{\theta}^{T} \mathbf{S}(\mathbf{e}') \right\},
\end{equation} 
\noindent where $\mathbf{S}(\mathbf{e})$ is the summary statistics that characterize the network structure such as number of edges, degree distribution, edge-wise shared partnership, etc. $K(\bs\theta)$ is the normalizing constant summed over all possible adjacency matrix $\mathbf{e}'$ and is often analytically intractable unless in small networks. Inference of $\bs\theta$ is often based on approaches with approximate $K(\bs{\theta})$, such as the Monte Carlo maximum likelihood estimation \citep{geyer1992constrained, hunter2006inference}. Eq \eqref{eq:ergm} is a simplified ERGM as we deal with CTN without nodal attributes in this study. In general, $\mathbf{S}$ may contain statistics not only constructed from $\mathbf{e}$ but also  nodal statistics for networks with nodal attributes.  

The steps of a general DP-ERGM procedure are as follows. Given a ERGM (either specified prior to the access to the observed data or chosen using a privacy-preserving procedure based on the observed by costing a portion of the total privacy budget), 1) derive the posterior distribution $\bs{\theta}$  given the likelihood function in Eq \eqref{eq:ergm} and a prior on $\bs{\theta}$; 2) obtain a sanitized  sample $\bs{\theta}^*$ from the posterior distribution with a pre-specified privacy budget $\epsilon$; 3) simulate  a network $\mathbf{e}^*$ via the ERGM parameterized by $\bs{\theta}^*$. If multiple sanitized networks are to be released, the above steps are repeated for $m>1$ times. 

In addition to DP-ERGM, we also examined a random response (RR) mechanism for perturbing edge information with DP guarantees  \citep{karwa2017sharing} and a debiased version of the RR mechanism \citep{DPERGM}. Both procedures perform significantly worse than the DP-ERGM procedure in the utility analysis performed in Section \ref{sec:study3} unless the privacy loss is high ($\epsilon>5$). The details on RR and RR-debias can be found in the supplementary materials. 
\subsection{Simulation Study}\label{sec:study3}
To evaluate statistical utility of sanitized CTNs, we conduct a simulation study. We simulated 500 sets of networks from an ERGM model with a single covariate $s$ (edge count). In each simulated network, there are 100 nodes. The networks were simulated to mimic real-life CTN (a CT dataset collected at the University of Notre Dame, USA, during the pandemic) in the degree distribution per individual. The real data are not shareable due to privacy and IRB reasons.  


The ERGM used in the DP-ERGM procedure contains  edge count as a single covariate. We applied an approach in \citet{liu2016model} to draw a privacy-preserving posterior sample on $\theta$  and also sanitized the edge count via the Laplace mechanism, which has a sensitivity of 1 (flipping a relation between two nodes changes the edge count in a network by at most 1). We equally split the total privacy budget $\epsilon$ between drawing a posterior sample of $\theta$  and sanitizing the edge count given a network. Given the privacy-preserving sample of $\theta$ and the sanitized edge count, we generated a privacy-preserving CTN under the constraint that its edge count equals to the sanitized edge count. We examine  $\epsilon\!=5, 2, 1,  0.5$.  The ERGM model fitting and network simulation were completed using R package \texttt{statnet}\citep{handcock2008statnet}. We conduct two utility analyses. In the first analysis,  we examine the preservation of qualitative information  and descriptive statistics in sanitized CTNs; in the second analysis, we run the ERGM on sanitized networks to examine the inference on the model parameter.  $m$ is set at 1 and 3, respectively, in these two analyses.

For the first utility analysis, we calculate some common network summary statistics, including edge counts, triangle counts,  degree distribution (DD),  and edgewise shared partners distribution (ESPD), and two node centrality measures in a sanitized network. Edge and triangle counts are the number of edges and triangles in a network.  The DD in a network  with $n$ nodes  consists of $d_k$ for $k\!=\!0,\ldots,n\!-\!1$, where $d_k$ is the number of nodes that share an edge with exactly $k$ other nodes. The ESPD consists of $\mbox{esp}_k/\mbox{edge count}$ for $k\!=\!1,\ldots,\le n(n-1)/2$, where $\mbox{esp}_k$ is the number of edges whose two nodes are both connected with exactly $k$ other nodes than themselves.  The betweenness centrality measures the centrality of a node in a graph and is defined for node $i$ as the proportion of the shortest paths  that connect nodes $j$ and $j'$ while passing through node $i$ ($j\ne j'\ne i)$ among all shortest paths that connect nodes $j$ and $j'$.  There are multiple definitions of closeness centrality and we use  $\big(\frac{A_i}{n-1}\big)^2/C_i$, where  $A_i$ is the number of reachable nodes from node $i$, and $C_i$ is the sum of distances from node $i$ to all reachable nodes. If no nodes are connected with node $i$, its closeness centrality is 0.  

\begin{figure}[!htb]
\centering\vspace{-9pt}
original \hspace{1cm} $\epsilon=0.5$ \hspace{1.2cm} $\epsilon=1$\hspace{1.5cm} $\epsilon=2$ \hspace{1.5cm} $\epsilon=5$ \\
\includegraphics[width=0.18\linewidth]{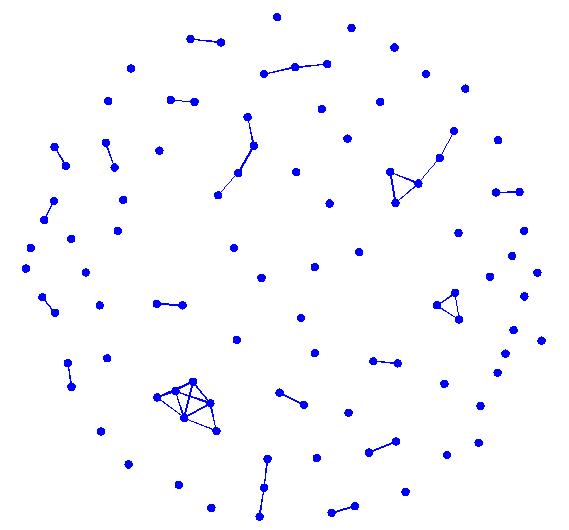}\hspace{2pt}
\includegraphics[width=0.18\linewidth]{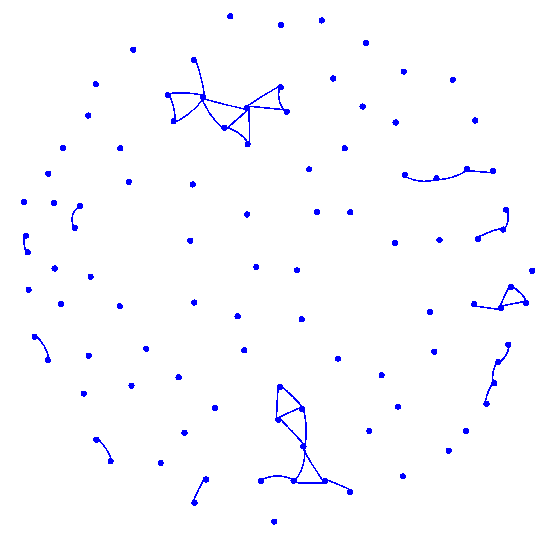}\hspace{2pt}
\includegraphics[width=0.18\linewidth]{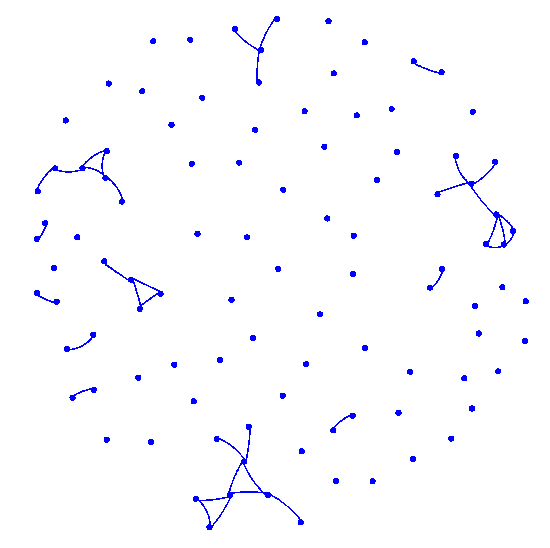}\hspace{2pt}
\includegraphics[width=0.18\linewidth]{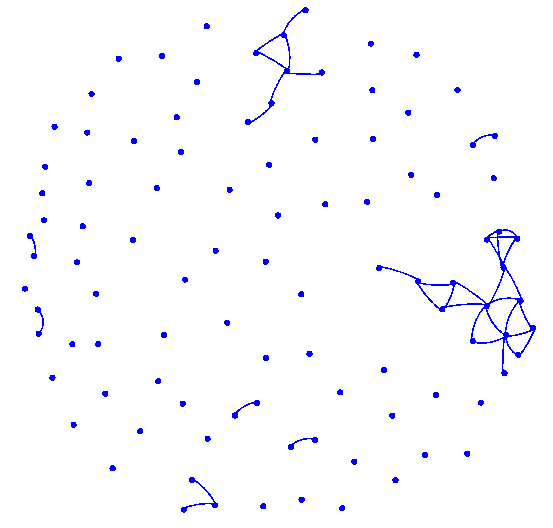}\hspace{2pt}
\includegraphics[width=0.18\linewidth]{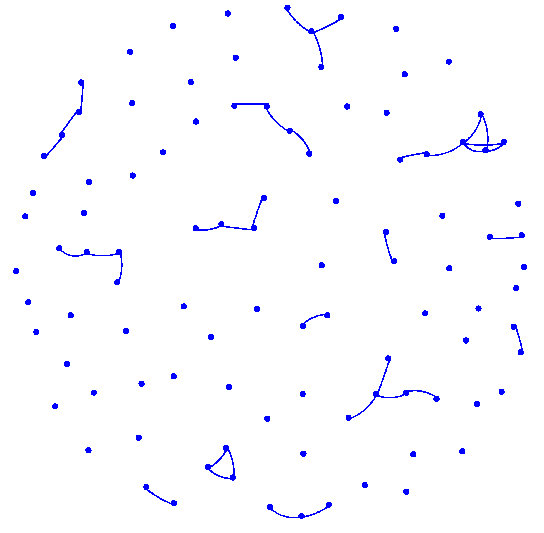}\\\vspace{-2pt}
{\small (a) examples of sanitized CTNs}\\\vspace{6pt}
\includegraphics[width=0.19\linewidth]{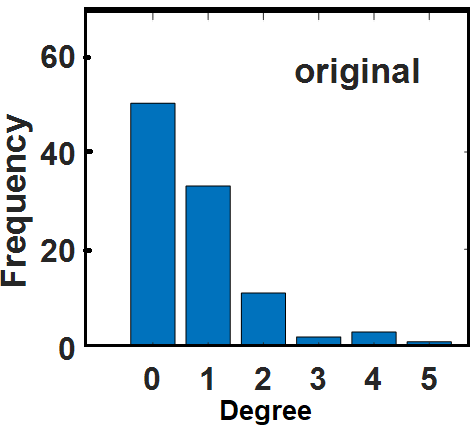}
\includegraphics[width=0.78\linewidth]{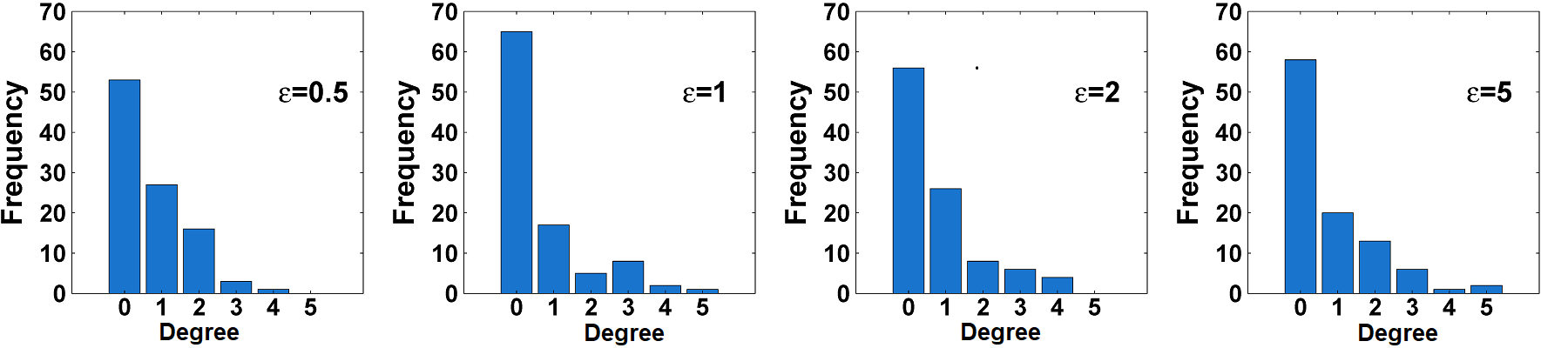}\\\vspace{-3pt}
{\small(b) degree distribution}\\
\includegraphics[width=0.3\linewidth]{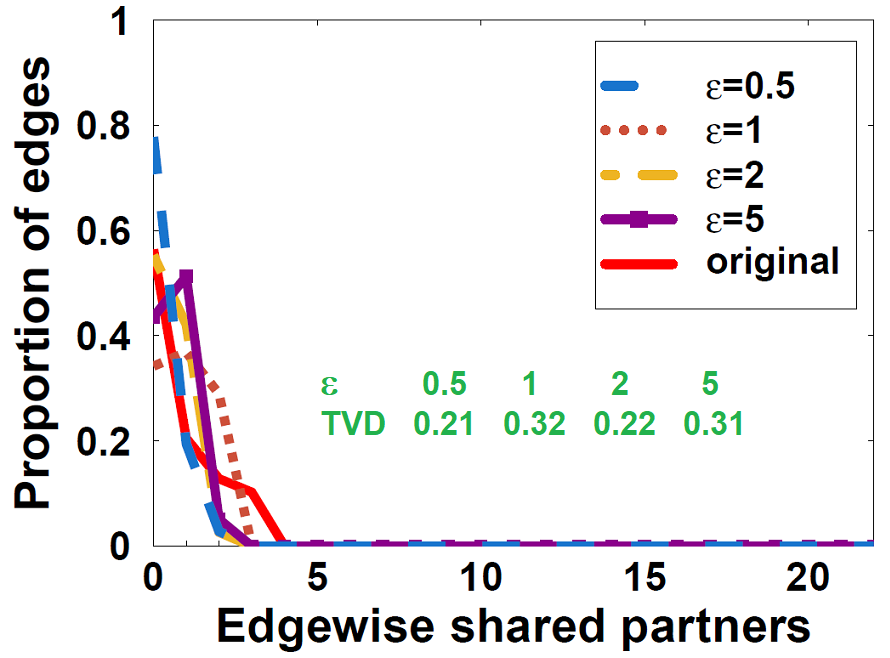}
\includegraphics[width=0.3\linewidth, trim= 0 0 1cm 1cm, clip]{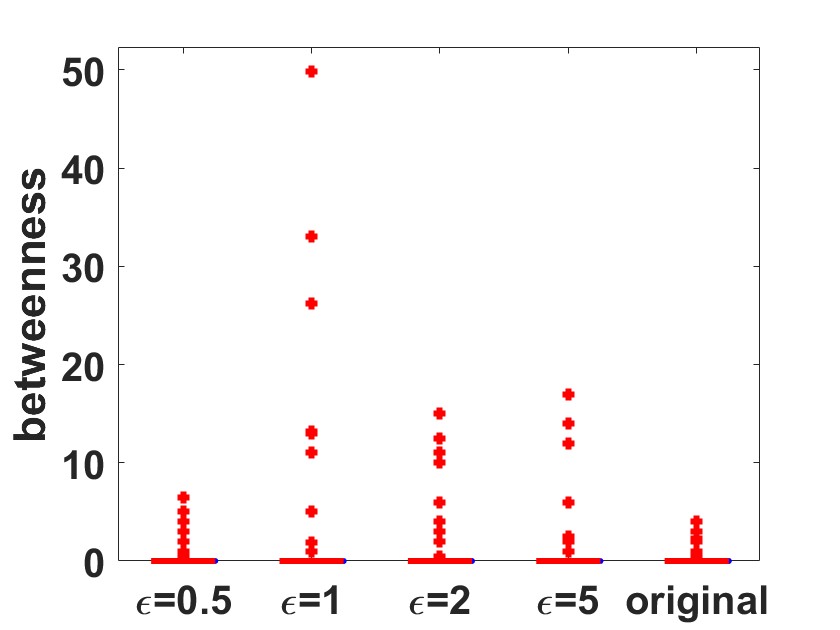}
\includegraphics[width=0.3\linewidth, trim= 6pt 0 1cm 0, clip]{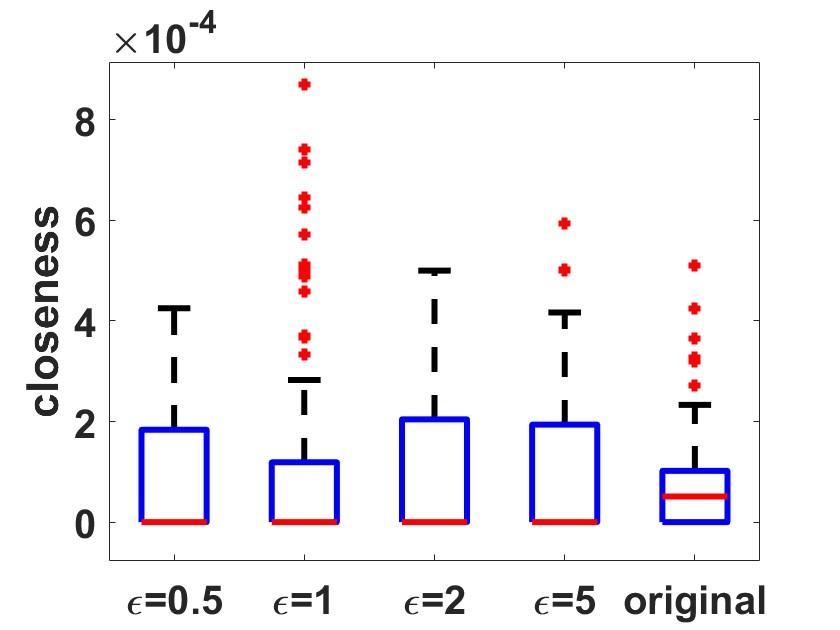}\\\vspace{-3pt}
{\small(c) Edgewise shared \hspace{0.8cm} (d) betweenness \hspace{1cm} (e) closeness}\\
{\small partner distribution \hspace{1.5cm} centrality  \hspace{1.7cm} centrality}\\ \vspace{-6pt}
\caption{Comparison between the original and sanitized CTNs via DP-ERGM on various network  structural statistics} \label{fig:stats}
\vspace{-6pt}
\end{figure}

The visualization of a single sanitized  CTN from one of the 500 repeats are presented in Figure \ref{fig:stats}(a) and provides a big-picture comparison between the sanitized vs the original networks in terms of density, clustering, etc. In summary, the density of the sanitized CTNs via DP-ERGM is similar to the original CTN  at all the examined $\epsilon$ values. Note the nodes in the sanitized networks do not  match the nodes in the original CTN as DP-ERGM samples a whole new surrogate network from a differentially private ERGM model for release. The edge and triangles counts of the original networks are 39 and 10, respectively. The average (standard deviation)  edge counts over  100 sanitized CTNs are 38 (6.4), 39 (3.1), 39 (1.4), and 39 (0.7)  at $\epsilon=0.5, 1, 2,$ and 5, respectively; the  average (standard deviation) triangle counts over  100 sanitized CTNs are 13 (9.2), 12 (7.4), 11 (6.8), and 11 (7.1) at $\epsilon=0.5, 1, 2,$ and 5, respectively. These numbers are consistent with  the observations in Figure \ref{fig:stats}(a).   Figures \ref{fig:stats}(b) and \ref{fig:stats}(c) depict the DD and  ESPD of the sanitized CTN. In the latter, we also calculate the total variance distance (TVD) in ESPD between the sanitized and original CTNs, which are presented in Figure \ref{fig:stats}(c). Figures \ref{fig:stats}(d) and \ref{fig:stats}(e) show the box plots of the  betweenness centrality and closeness centrality of the 100 nodes in the original and sanitized CTNs.  Though there is some deviation in he DD, ESPD, and the distributions of the centrality measures in the sanitized CTNs from the original, the deviation is rather mild. In addition, the statistics are relatively stable across $\epsilon$. 

For the second utility analysis,  we fitted the ERGM on the sanitized CTNs  to obtain privacy-preserving inference on $\theta$, the coefficient associated with edge count in ERGM, via the inferential rule in Eqs \eqref{eqn:T} and \eqref{eqn:inference}. The results are presented in Table \ref{tab:ergm}. In summary, the results are acceptable for the ERGM analysis at all examined $\epsilon$ (especially for CP).
\begin{table}[!htb]\vspace{-4pt}
\caption{Inference of ERGM parameter based on sanitized CTNs ($m=3$, 500 repeats)} \label{tab:ergm}\vspace{-5pt}
\centering
\resizebox{0.55\linewidth}{!}{
\begin{tabular}{c cccccccc}
\hline
 & original$^\dagger$ & $\epsilon=5$ & $\epsilon=2$ & $\epsilon=1$ &$\epsilon=0.5$\\
\hline
bias & -0.021 & -0.021 & -0.026 & -0.031 & -0.051 \\
RMSE & 0.171& 0.172 & 0.174 & 0.187 & 0.260\\
CP & 0.942 &0.954 & 0.954 & 0.952 & 0.944\\
\hline
\end{tabular}}
\vspace{-12pt}
\end{table}

\subsection{Summary}\vspace{-3pt}
The simulation study suggests that the DP-ERGM approach can produce privacy-preserving CTNs that are structurally similar to original CTNs by various statistical measures. In addition, the utility of sanitized CTNs is relatively insensitive to $\epsilon$ for the examined range of $[0.5,5]$, implying that a small $\epsilon$ can be used to provide strong privacy guarantees without sacrificing much of the utility. The sanitized CTNs can be shared with researchers who are interested in learning more about CTNs during the pandemic, without compromising individual privacy at a pre-specified privacy cost. 

\section{Discussion}\label{sec:discussion}
We use three common data types -- surveillance case numbers, case location information, and contact tracing networks -- collected during the COVID-19 pandemic to demonstrate the release and sharing privacy-preserving data.
In each data case, we apply randomized mechanisms  with formal privacy guarantees to sanitize and release information aiming at preservation of statistical utility and aggregate information that can be used to infer underlying population parameters, as shown in the simulation studies and real-life applications. The approaches do not target at learning individual-level information, which not only conflicts with the goal of privacy protection, but is also unnecessary for the purposes of mining and understanding the population-level information. 

DP and its various extensions are state-of-the-art concepts in privacy research and are quickly adopted in practice. Some of the methods we have demonstrated in the study are basic and have been routinely  applied for privacy protection, such as the flat sanitizer; and some are recently proposed, such as DP-ERGM. For all the data types and examples examined in this study, synthetic data are generated and released at a pre-specified privacy budget and users may perform  their own analysis on the synthetic data without having to worry about additional privacy loss. Our simulation studies suggest that different DP methods may lead to different utility and vary in the easiness of implementation  for a given statistical analysis procedure at the same privacy cost, an observation that is well documented in the literature and also one of the reasons why new DP methods are constantly proposed to improve on the existing methods with either better utility or more straightforward implementation. In addition, absolute privacy protection for individuals in a dataset only exists on paper unless the released information is completely random or independent of the dataset. In reality,  there is always some loss in privacy when releasing new and useful information; larger privacy loss parameters means sacrifice in privacy in hope for better utility in released information.  Choice of a proper privacy loss is a key step when implementing DP procedures.

We hope our study and the examples shed light on privacy-preserving sharing of COVID-19 data to help promote and encourage more data sharing for research  use. For future work on this topic, we will continue to develop methods to deal with more complicated COVID-19 data sharing situations, such as releasing travel trajectories of COVID-19 patients, longitudinal data, and dynamic CTNs, CTNs with nodal attributes, among others.


\setlength{\bibsep}{3pt plus 3pt}
\bibliographystyle{plainnat}
\bibliography{ref.bib}

 
\clearpage
\setcounter{page}{1}
\setcounter{figure}{0}
\setcounter{table}{0}
\setcounter{equation}{0}
\setcounter{section}{0}
\renewcommand{\thepage}{S\arabic{page}}
\renewcommand{\thesection}{S\arabic{section}}
\renewcommand{\thetable}{S\arabic{table}}
\renewcommand{\thefigure}{S\arabic{figure}}
\renewcommand{\theequation}{S\arabic{equation}}

\begin{center} \Large\bf{Supplementary Materials to ``\emph{Some Examples of Privacy-preserving Publication and Sharing of COVID-19 Pandemic Data}''} \end{center}

\normalsize
\subsection*{UH and UHp for Surveillance Case Number Release}
\begin{figure}[!htb]
\vspace{-9pt}\centering
\includegraphics[width=0.6\textwidth]{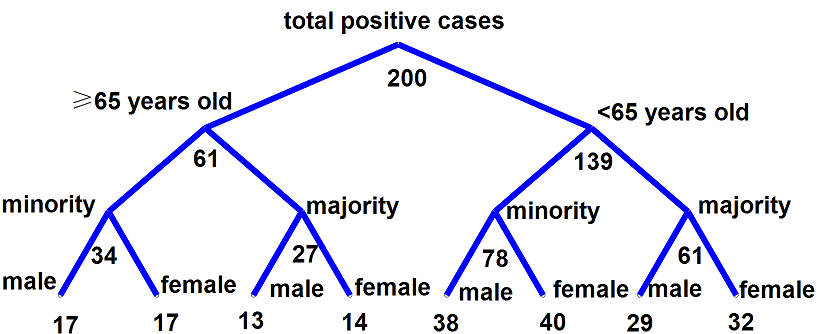}
\caption{A count hierarchical tree with three binary attributes }\label{fig:tree} \vspace{-12pt}
\end{figure}

The UH approach forms a hierarchical tree among the data attributes and injects noise to each the node count in each layer of the tree, explores equality constraints between each parent node and its children nodes in the tree to improve the accuracy of the sanitized count of the parent nodes (low-order marginals) and release the final corrected counts from the whole tree.  Figure \ref{fig:tree} displays a 4-layer hierarchical tree formed in the UH approach on a data set with 3 variables (age group, minority/majority, sex). We refer to the node at the top of the tree as the root  (layer 1) and those at the bottom as the leaf nodes (layer 4). The age nodes at layer 3 are parents to the race/ethnicity nodes in layer 3, which are the parent nodes to the sex nodes in layer 4.  There is no particular ordering among the three attributes in the example in Figure \ref{fig:tree}. We can place the attributes in the middle layers of the trees that would enjoy a lower mean squared error (MSE) (MSE for a sanitized count $\tilde{x}$ is $ \mathbb{E}_{\M}(\tilde{x}-x)^2$, where $x$ is the original count and  the expectation is taken over the distribution of the randomized algorithm). in their marginal sanitized counts relative to their original counts, compared to the MSE resulting from a simple sum of the directly sanitized counts of the most granular cells as done in the flat sanitizer.

The UH procedure is implemented in 3 steps. First, since each layer is sanitized, the total budget $\epsilon$ should be split among the layers following the sequential composition principle in DP  \citep{mcsherry2007mechanism}. For illustration purposes, we assume each layer receives $1/l$ of the total $\epsilon$, where $l$ is the height of the tree (other privacy allocation schemes across the layers can also be used). and  $l=4$ in Example 1. The count $h[v]$ in each node $v$ in the tree is sanitized via the Laplace mechanism Lap$(0,l\epsilon^{-1})$; that is, $\widetilde{h}[v]=h[v]+e, \mbox{where $e\sim$ Lap} (l\epsilon^{-1})$,  where  $\widetilde{h}[v]$ is the sanitized count. In step 2, intermediate node count $z[v]$ for each node $v$ is obtained via Eq (\ref{eqn:z}),
\begin{align}\label{eqn:z}
\!z[v]\!=\!
\begin{cases}
\widetilde{h} [v], \mbox{ if $v$ is the leaf node}\\
\frac{k^{l}-k^{l-1}}{k^{l}-1}\widetilde{h}[v]+ \frac{k^{l-1}-1}{k^{l}-1}\sum_{u \in succ(v)} z[u],\mbox{ o.w.}
\end{cases},
\end{align}
where $\mbox{succ}(v)$  denotes the set of children nodes to parent node $v$ and $k$ is the number of children per parent node, which is assumed to be the same for each parent ($k=2$ in example 1). The reason behind Eq \eqref{eqn:z} is that for the nodes not from the bottom layer (the non-leaf nodes), a sanitized count comes from two sources (the node being sanitized, and the summation from its children nodes) so Eq \eqref{eqn:z} calculates a weighted average of the two. Obviously, $z[v]$  may no longer equal to the sum of the node counts of its children nodes, violating the equality constraints in contingency tables. This inconsistency is corrected via  Eq \eqref{eqn:h}, yielding the final sanitized count $h^*[v]$
\begin{align}\label{eqn:h}
h^*[v]\!=\!
\begin{cases}
z[v] \mbox{ if $v$ is the root node}\\
z[v]+k^{-1}\left(h^*[u]-\sum_{w \in succ(u)} z[w]\right), \mbox{ o.w.}
\end{cases},
\end{align}
where $u$ is the parent mode to node $v$, $\mbox{succ}(u)$  contains the children nodes to parent node $u$,  and $h[u]-\sum_{w \in succ(u)} z[w]$ is the correction term to ensure the  equality constraint holds for each parent node in the tree.

We extend the UH approach to sanitizing a proportion tree (Figure \ref{fig:tree}) in place of a count tree and name it the UHp approach (``p'' in the name ``UHp'' stands for ``porportion''), in cases where the total sample size $n$ is public information and can be released directly, or when it is desirable not to alter $n$ from a statistical inferential perspective as $n$ is critical for inferences such as inferential efficiency.  
\begin{figure}[!htb]
\vspace{-9pt}\centering
\includegraphics[width=0.6\textwidth]{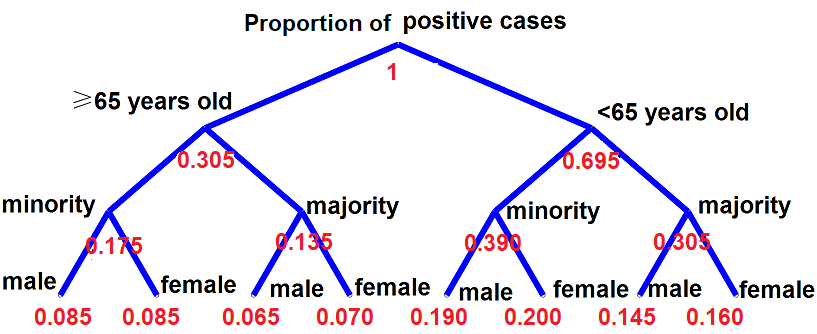}
\caption{A proportion hierarchical tree with three binary attributes }\label{fig:treep} \vspace{-12pt}
\end{figure}

The sanitization process for UHp is similar to UH with a few modifications. First, given the proportion at the top layer is always 1, there is no need to sanitize the node and the total $\epsilon$ is only needed to split into $l-1$ layers. Second, the Laplace distribution from which the noise is drawn becomes Laplace($0,(l-1)\epsilon^{-1}/n$) as the global sensitivity for proportion is $1/n$. Third, after obtaining $\widetilde{h}[v]=h[v]+e$, where $e\sim$ Lap $(l-1)\epsilon^{-1}/n)$ for all the non-root node proportions, we  normalize the proportions in layer 2 as in $\widetilde{h}[v^{(2)}]=\widetilde{h}[v^{(2)}]/\sum_{u}\widetilde{h}[u]$, where $u$ refers to all the nodes in layer 2, so that the layer-2 proportions sum up to 1, honoring the constraint of $\tilde{h}[v]=h[v]=1$ for the root node. The steps in Eqs \eqref{eqn:z} and \eqref{eqn:h} after the normalization step remain the same as in the UH approach.  After the sanitized proportions are obtained, the corresponding counts can be obtained by multiplying the proportions with the total $n$.

Similar to the flat sanitizer, the sanitized counts or proportions in  the UH and the UHp approaches can be negative as the support of the Laplace distribution is $\mathbb{R}$. In addition, the sanitized proportions may be $>1$. We applied the same methods as used for the flat sanitizer to deal with negative counts and in the case of a fixed upper bound such as the proportions add up to 1 and when the total count is fixed.

\subsection*{Simulation study and CDC death count application for UH and UHp}
\begin{figure}[!htb]
\vspace{-9pt}\centering
\includegraphics[width=1\textwidth]{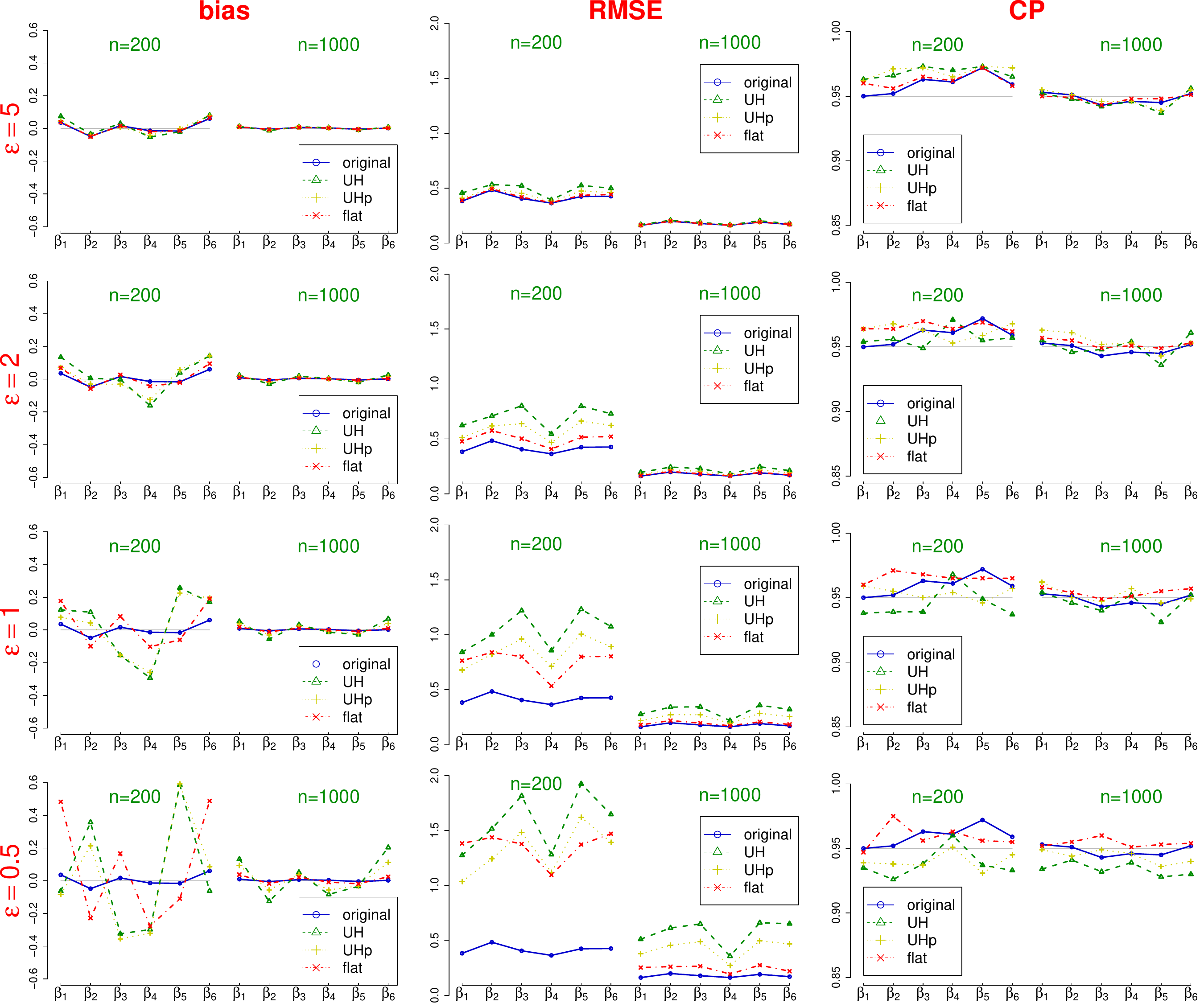}
\caption{Privacy-preserving inference of the log-linear model  based on sanitized counts by different methods in the simulation study ($m=3$, 500 repeats)} \label{fig:sim1UHUHp}\vspace{-9pt}
\end{figure}

In the simulation study, for both UH and UHp,  the tree height is $l=4$ as  there are 3 attributes -- $X_1$ is layer 2, $X_2$ in layer 3, and $X_3$ in layer 4 -- and $k=2$ as all attributes are binary. The  simulation results are presented in Figure \ref{fig:sim1UHUHp}, together with the flat Laplace sanitizer and the original results for comparison.

\begin{figure}[!htb]
\vspace{-3pt}\centering
\includegraphics[width=1\textwidth]{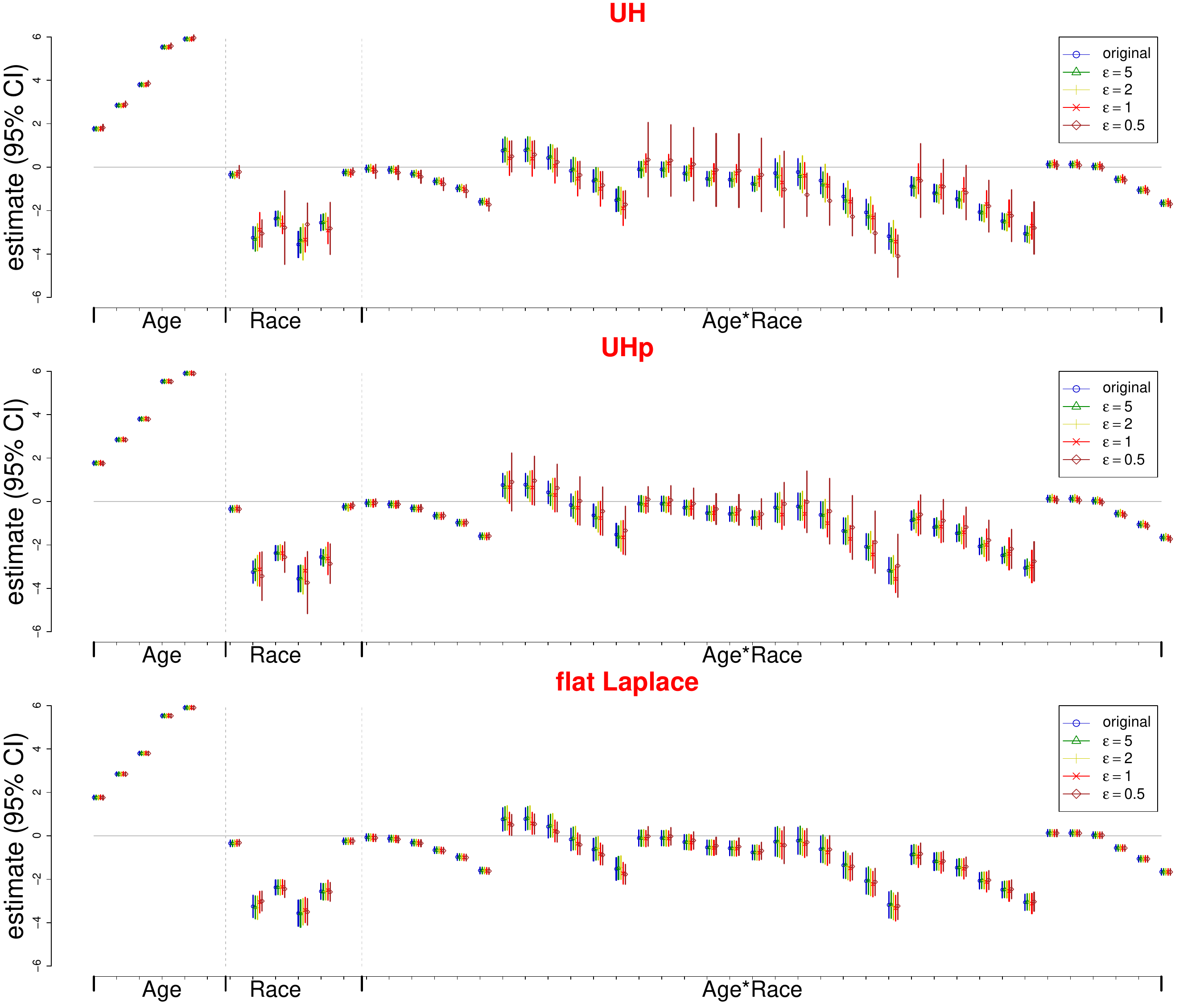}
\caption{Privacy-Preserving results from the Log-linear model fitted on the CDC COVID-19 death data }\label{fig:cdc3}
\end{figure}

For the application to the CDC COVID-19 death count data, $l=3$ and $k=7$ in the hierarchical tree for both UH and UHp. We placed age in layer 2 and race/ethnicity in layer 3 and don't expect the ordering would affect the results of the analysis we conducted in a statistically meaningful way. The results are presented in Figure \ref{fig:cdc3}, together with the flat Laplace sanitizer and the original results for comparison.

\clearpage
\subsection*{Examples of sanitized  US COVID-19 death counts by the Flat Laplace sanitizer ($m=3$ and $\epsilon=0.5$)}\vspace{-6pt}
\begin{table}[!htb]
\centering 
\caption{Three sets of sanitized U.S. COVID-19 death counts by age group and race/ethnicity on May 24, 2022 ($m=3,\epsilon=0.5$) by the flat Laplace sanitizer}\label{tab:death.m3}
\vspace{-3pt}
\resizebox{1\textwidth}{!}{
\begin{tabular}{lcccccccc} 
\hline
Age  (ys) & \multicolumn{7}{c}{Race/Ethnicity}\\
\cline{2-8}
group  & NH White &NH Black &NH AIAN &NH Asian &NH NHPI &NH Mix &Hispanic & Total\\
\hline
<17 &385 &273 &19 &38 &14 &25 &302 &1056 \\
18-29 &2265 &1480 &183 &181 &33 &78 &2018 &6238 \\
30-39 &6665 &4140 &570 &561 &151 &160 &5916 &18162 \\
40-49 &17277 &8939 &1024 &1200 &267 &313 &13979 &43000 \\
50-64 &97407 &35752 &3196 &5310 &703 &955 &43655 &186979 \\
65-74 &141417 &37760 &2913 &7431 &498 &914 &38416 &229350 \\
>75 &380635 &54588 &3210 &16515 &447 &1383 &56700 &513478 \\
Total &646051 &142933 &11115 &31236 &2114 &3827 &160986 &998262 \\
\hline
\end{tabular}}
\resizebox{1\textwidth}{!}{
\begin{tabular}{lcccccccc} 
\hline
Age  (ys) & \multicolumn{7}{c}{Race/Ethnicity}\\
\cline{2-8}
group  & NH White &NH Black &NH AIAN &NH Asian &NH NHPI &NH Mix &Hispanic & Total\\
\hline
<17 &387 &289 &19 &34 &11 &35 &304 &1079 \\
18-29 &2273 &1492 &183 &197 &54 &86 &2014 &6299 \\
30-39 &6660 &4124 &568 &571 &147 &157 &5917 &18145 \\
40-49 &17269 &8930 &1029 &1216 &279 &314 &13972 &43010 \\
50-64 &97386 &35756 &3200 &5315 &723 &953 &43647 &186982 \\
65-74 &141407 &37753 &2890 &7426 &513 &918 &38420 &229327 \\
>75 &380591 &54568 &3205 &16518 &450 &1382 &56706 &513420 \\
Total &645974 &142912 &11095 &31278 &2177 &3846 &160980 &998262 \\
\hline
\end{tabular}}
\resizebox{1\textwidth}{!}{
\begin{tabular}{lcccccccc} 
\hline
Age  (ys) & \multicolumn{7}{c}{Race/Ethnicity}\\
\cline{2-8}
group  & NH White &NH Black &NH AIAN &NH Asian &NH NHPI &NH Mix &Hispanic & Total\\
\hline
<17 &392 &284 &19 &29 &10 &29 &309 &1072 \\
18-29 &2236 &1507 &189 &196 &48 &62 &2007 &6243 \\
30-39 &6659 &4146 &563 &562 &150 &153 &5903 &18135 \\
40-49 &17260 &8933 &1002 &1208 &287 &316 &13983 &42987 \\
50-64 &97418 &35743 &3198 &5312 &719 &964 &43674 &187027 \\
65-74 &141398 &37767 &2897 &7437 &516 &903 &38420 &229339 \\
>75 &380604 &54573 &3192 &16513 &460 &1391 &56724 &513458 \\
Total &645966 &142953 &11060 &31255 &2190 &3818 &161020 &998262 \\
\hline
\end{tabular}}
\resizebox{1\textwidth}{!}{
\begin{tabular}{l}
Race/ethnicity = 'unknown' is not included in the table.\\
NH = Non-Hispanic; AIAN = American Indian or Alaska Native; NHPI = Native Hawaiian or Other Pacific Islander; \\  
"Mix" means "more than one race"\\
\hline
\end{tabular}}
\end{table}

\clearpage
\subsection*{The RR mechanism and the RR-debiased procedure}
The RR mechanism for sanitizing edges in a network works as follows. Let $p_{ij}$ denote the probability  the original edge $e_{ij}=1$ is retained  and $q_{ij}$ be the probability that $e_{ij}=0$ is retained after santization for nodes $i\ne j=1,\ldots,n$. To satisfy $\epsilon_{ij}$-DP ($\epsilon_{ij}$ edge DP precisely speaking; see \citet{DPERGM} for details) in the sanitization of the relational information $e_{ij}=1$ , one may set $p_{ij}=q_{ij}=e^{\epsilon_{ij}}/(1+e^{\epsilon_{ij}})$. When there is no particular reasons for using different $\epsilon_{ij}$ for different pairs of nodes, one may set $\epsilon_{ij}\equiv\epsilon$ and the probability of edge flipping in the network is 
\begin{equation}\label{eqn:RR}
  p_{ij}=q_{ij}\equiv 1/(1+e^{\epsilon}). 
\end{equation}
If all edges are mutually independent, the total cost for sanitizing the whole network is also $\epsilon$ per the parallel composition principle. 


\citet{DPERGM} employs a debiasing approach as an attempt to remove bias in sanitized networks via the RR mechanism (with edges $e_{ij}^*$) by synthesizing new networks  with edges $\tilde{e}^*_{ij}$  given a RR-sanitized network. Specifically, 
\begin{align}
\tilde{e}^*_{ij}|e_{ij}^*=1&\sim\mbox{Bern}(p_1), 
\mbox{ where } p_1 = \frac{(p+q-1)q}{(2q-1)p},\label{eqn:p1}\\
\tilde{e}^*_{ij}|e_{ij}^*&\sim\mbox{Bern}(p_0), 
\mbox{ where } p_0 = \frac{q(p+q-1)}{(1-p)(2q-1)},\label{eqn:p0}
\end{align}
where  $q=e^{\epsilon}/(1+e^{\epsilon})$ is the probability of retaining an original edge by RR and $p$ is the proportion of all $e^*_{ij}=1$ in row $i$ of the adjacency matrix of the synthetic network generated by RR (without the diagonal element), and.  Synthetic networks via RR-debiased can be summarized and analyzed in the same way as the original network including  descriptive statistics, visualization, and inference.  For inference, there is no need to explicitly model the RR mechanism or the subsequent debiasing/sanitization process if $m>1$  sets of synthetic networks are released. The debiasing procedure does not use the information from the original network and thus maintains the privacy guarantees, but at the cost of introducing another layer of variability. The debiased sanitized network is made of edges $Y^*$ drawn from two Bernoulli distributions, depending on whether the synthetic edge $Y'$ from the DWRR is 1 or 0.

\subsection*{Simulation Study on RR and RR-debiased}
For RR, the probability of flipping an edge per Eq \eqref{eqn:RR} is $(1+e^5)^{-1}=0.7\%,(1+e^2)^{-1}=11.9\%, (1+e)^{-1}=26.9\%$ and $(1+e^{0.5})^{-1}=37.5\%$ at $\epsilon\!=5, 2, 1,  0.5$, respectively. Though the probability retaining the original relation between nodes $i$ and $j$ is very low at $\epsilon=5$, but the number of edges is expected to double ($39e^5/(1+e^5)+(4950-39)e^5/(1+e^5)=71.6$ where 39 is the edge count in the original network). 

\begin{figure}[!htb]\centering
\begin{minipage}{0.22\textwidth}\centering
\includegraphics[width=0.8\linewidth]{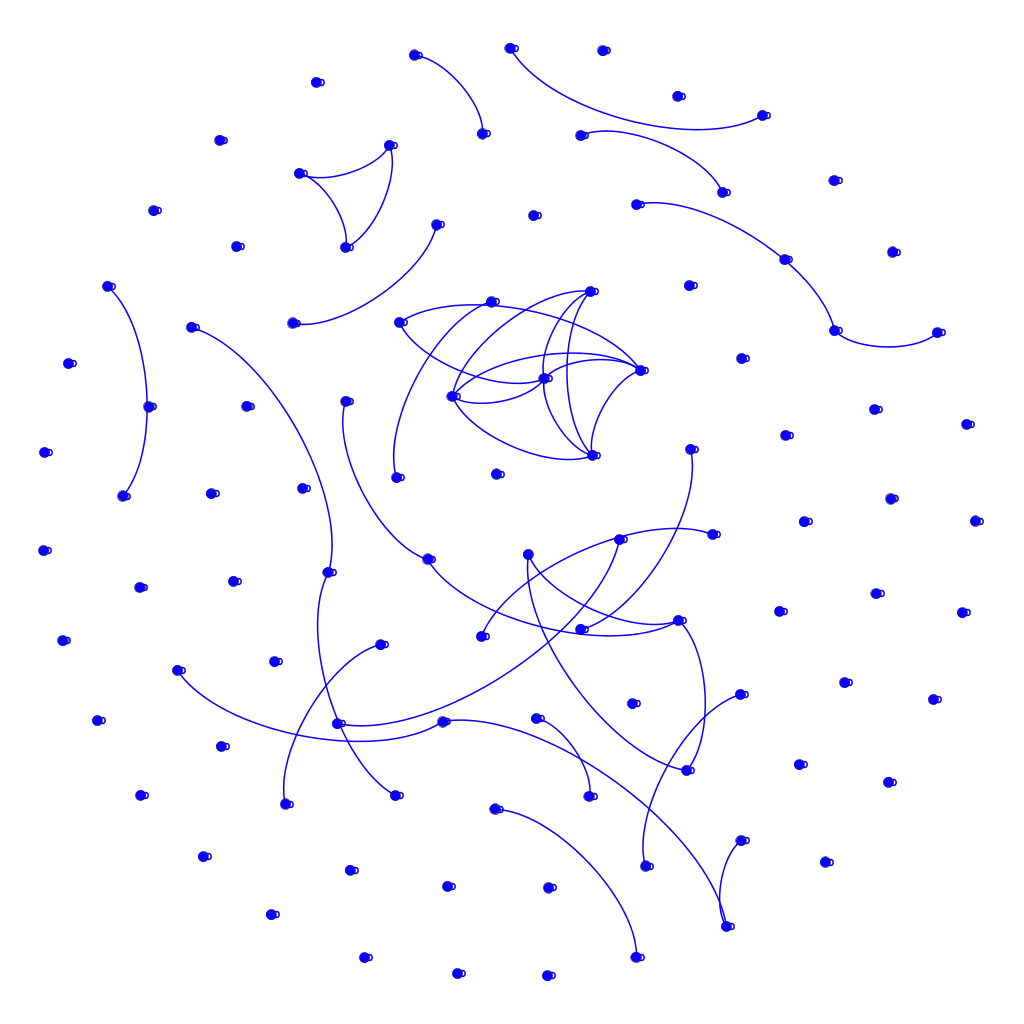}\\
(a) original
\end{minipage}
\begin{minipage}{0.75\textwidth}\centering
\includegraphics[width=0.23\linewidth, trim=1cm 1cm 1cm 6pt,clip]{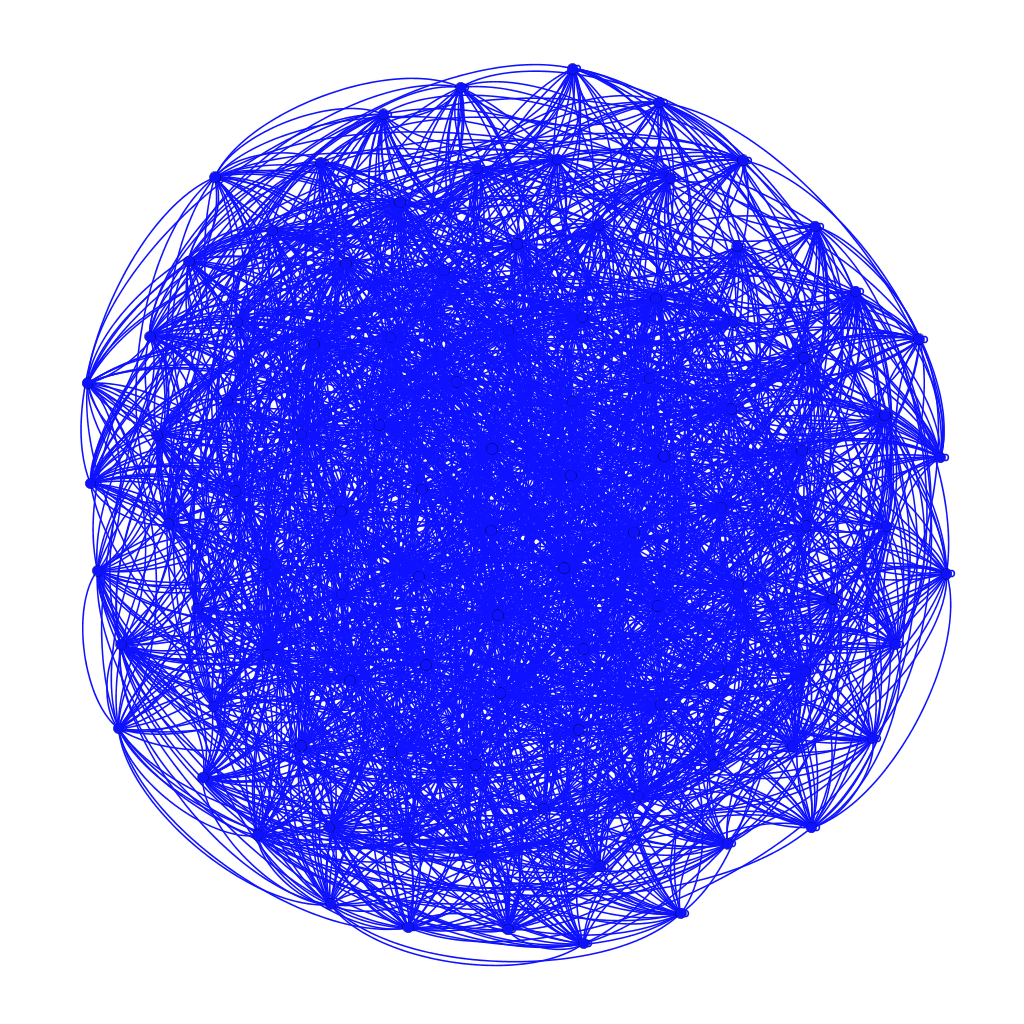}\hspace{3pt}
\includegraphics[width=0.23\linewidth, trim=1cm 1cm 1cm 6pt,clip]{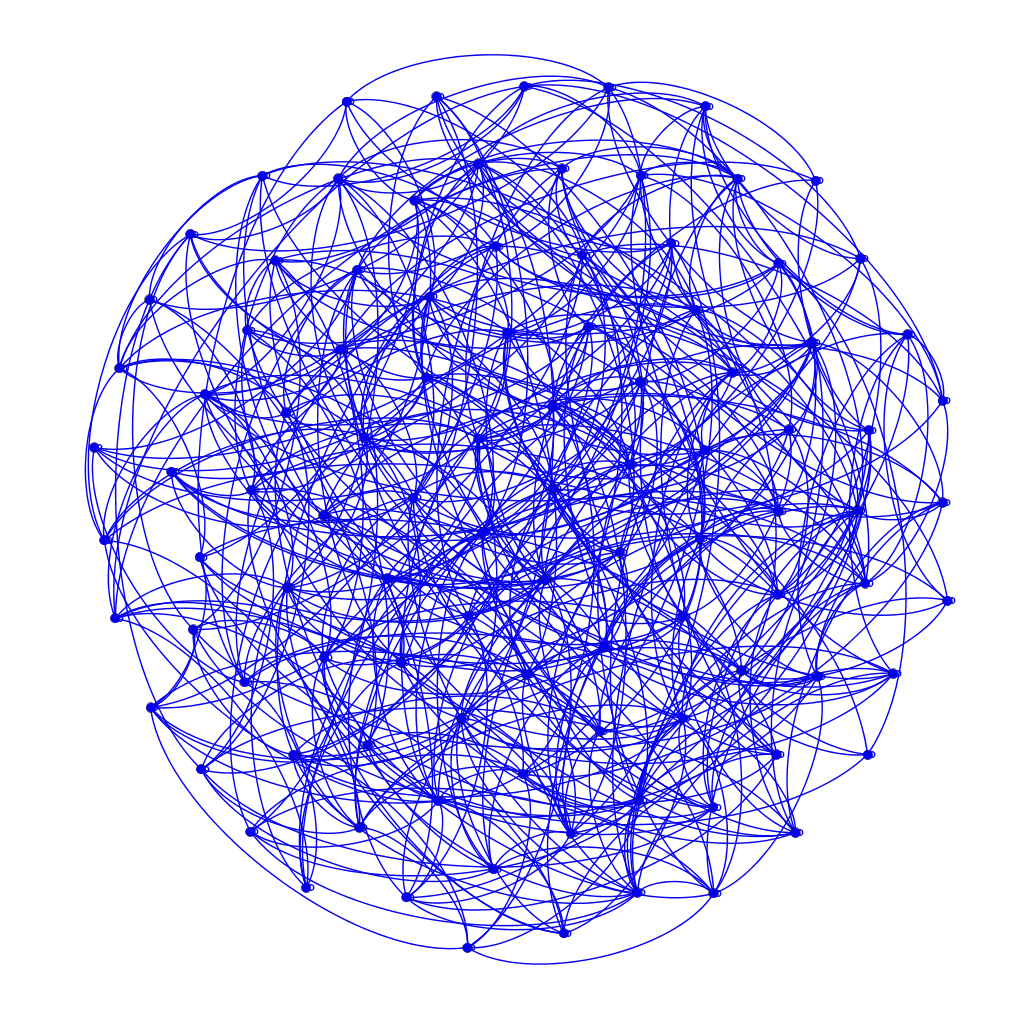}\hspace{3pt}
\includegraphics[width=0.23\linewidth, trim=1cm 1cm 1cm 6pt,clip]{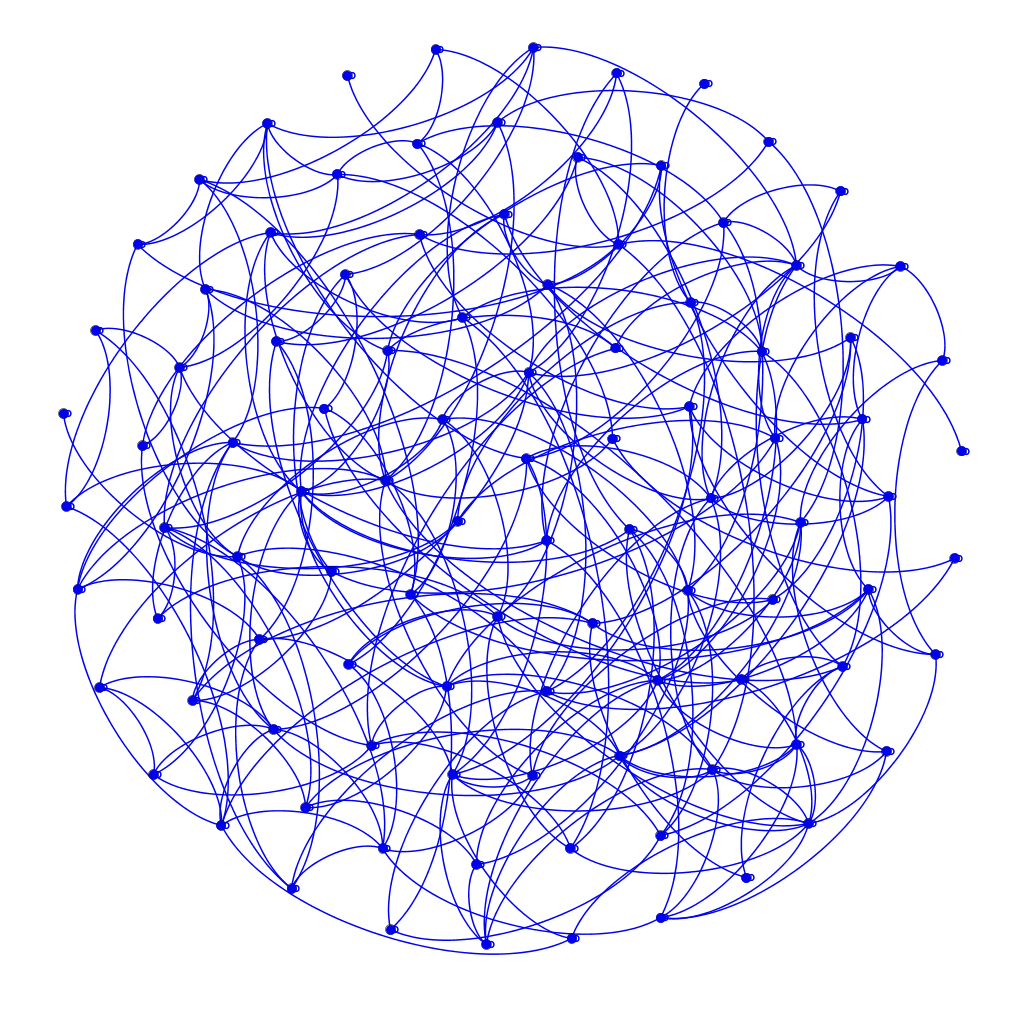}\hspace{3pt}
\includegraphics[width=0.23\linewidth, trim=1cm 1cm 1cm 6pt,clip]{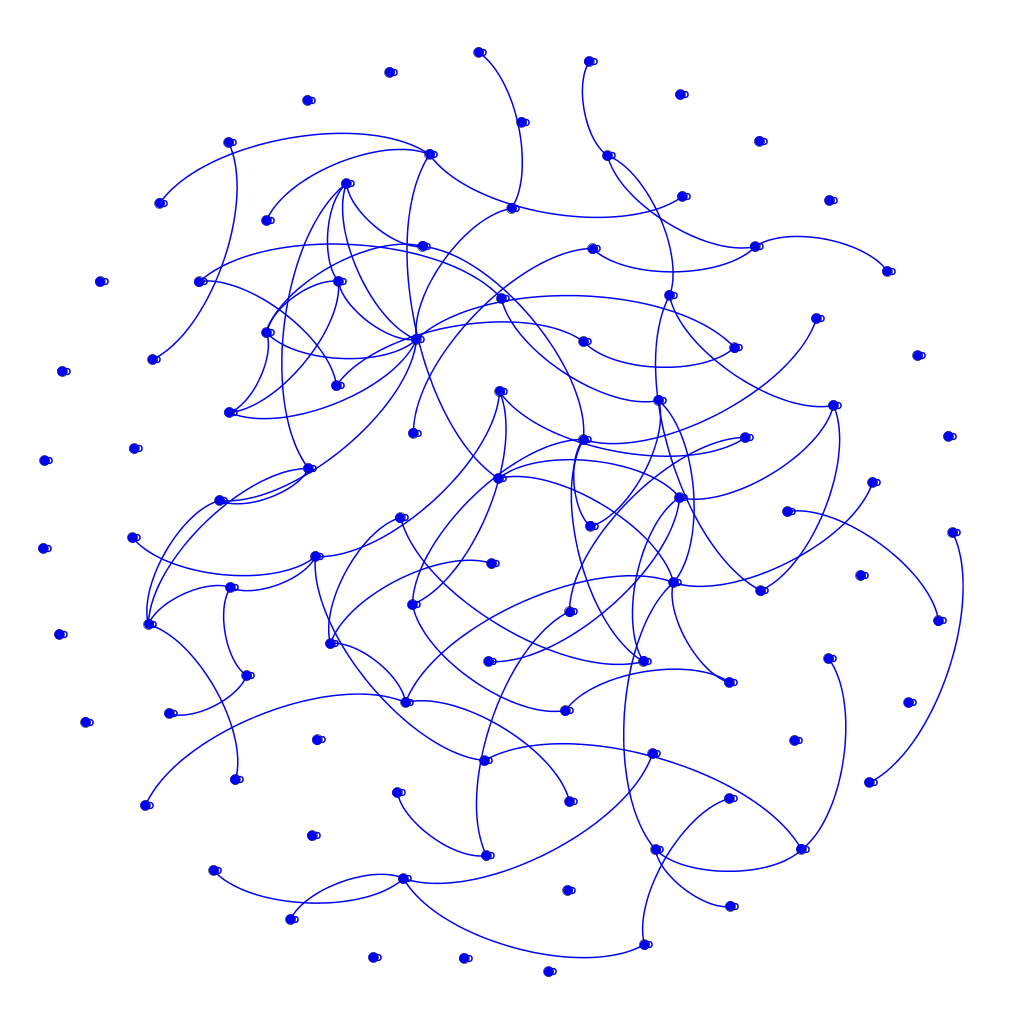}\\
(c) RR\vspace{-3pt}\\
\includegraphics[width=0.23\linewidth, trim=1cm 1cm 1cm 6pt,clip]{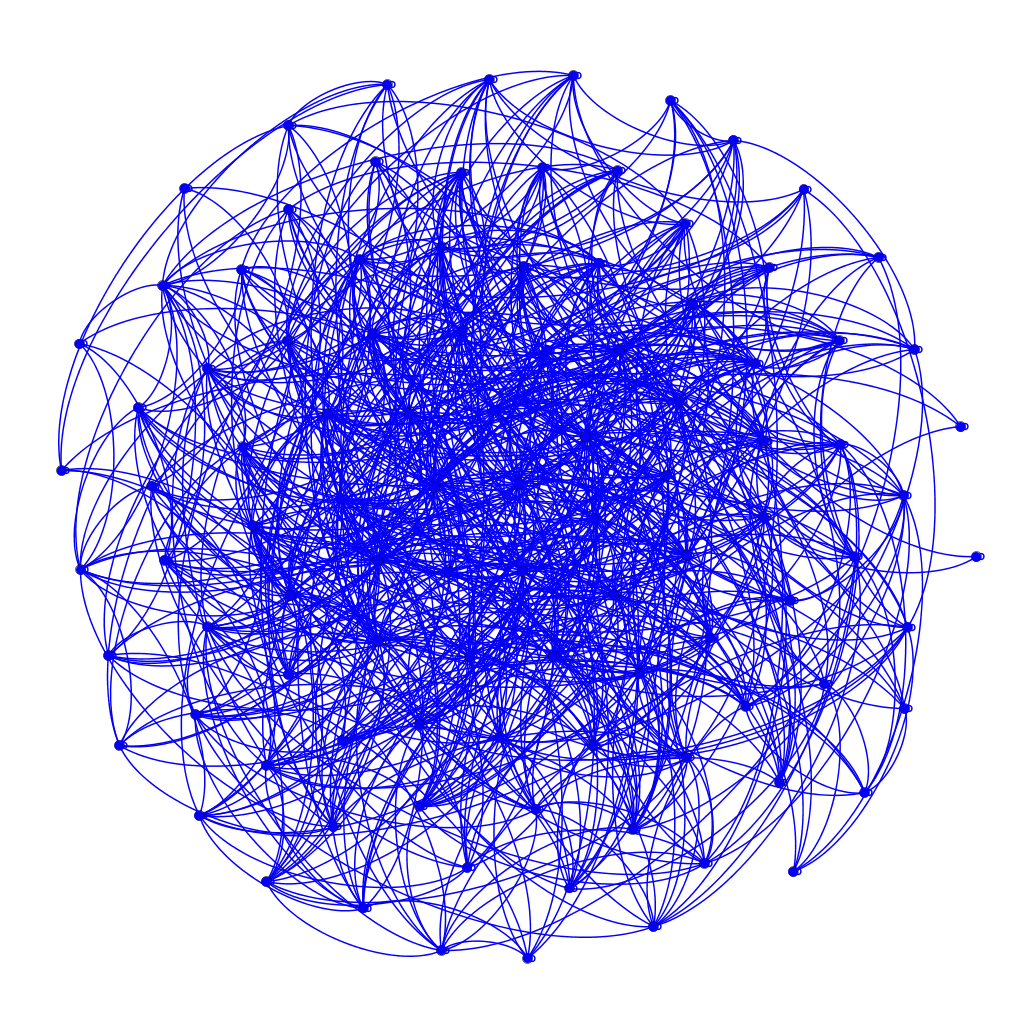}\hspace{3pt}
\includegraphics[width=0.23\linewidth, trim=1cm 1cm 1cm 6pt,clip]{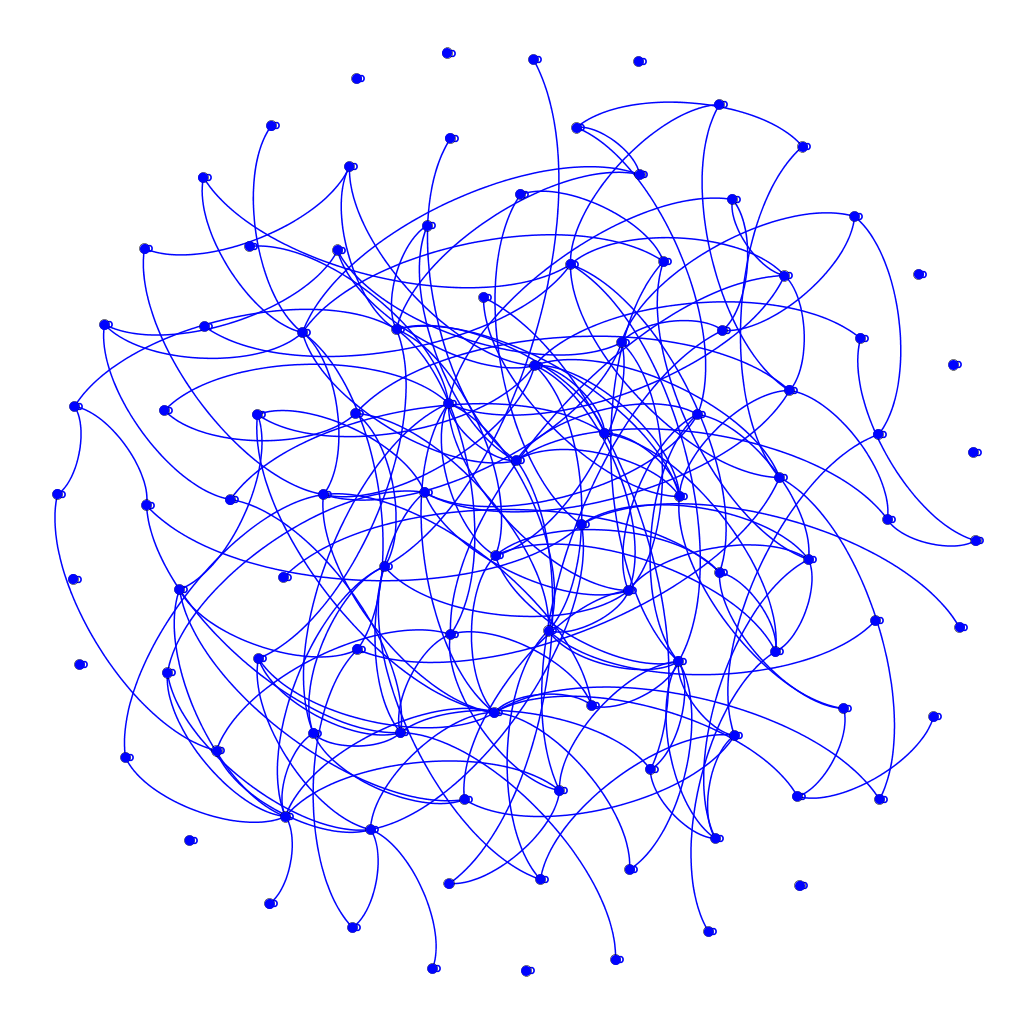}\hspace{3pt}
\includegraphics[width=0.23\linewidth, trim=1cm 1cm 1cm 6pt,clip]{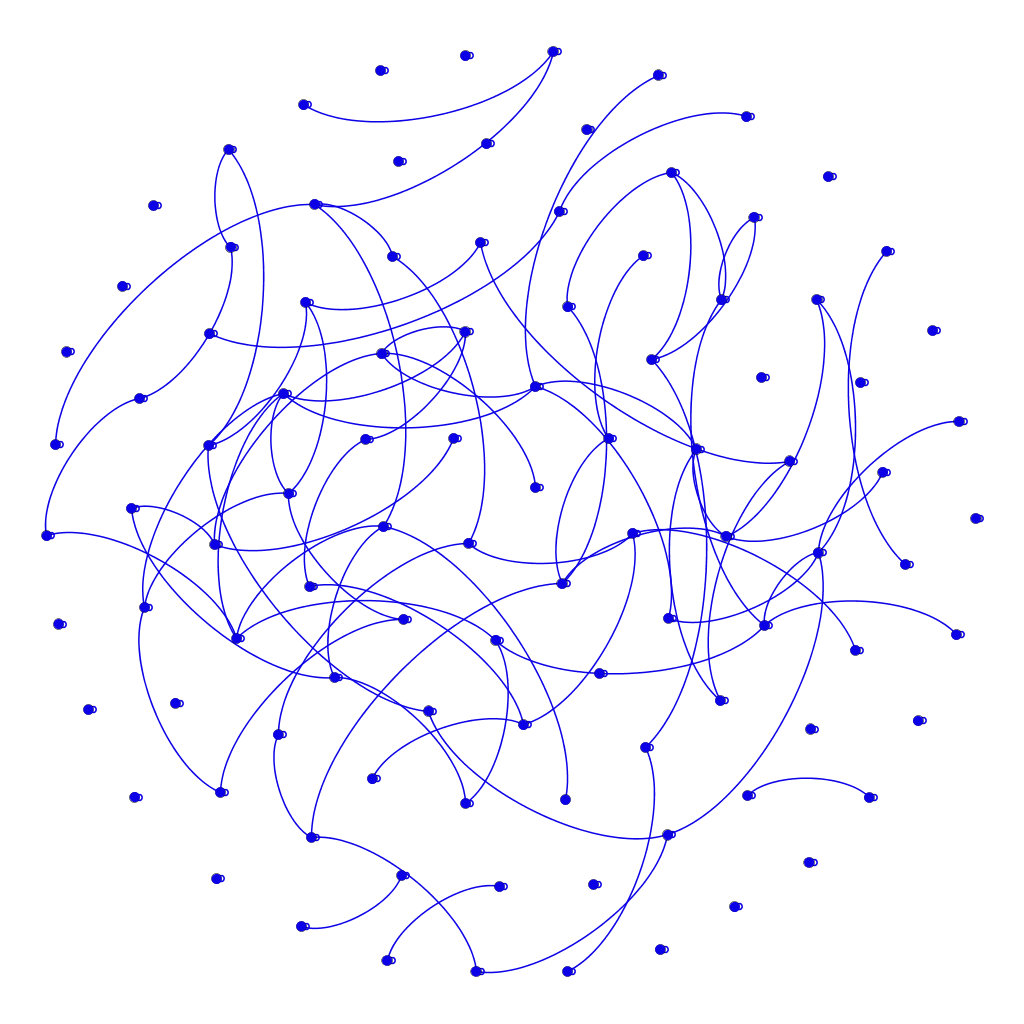}\hspace{3pt}
\includegraphics[width=0.23\linewidth, trim=1cm 1cm 1cm 6pt,clip]{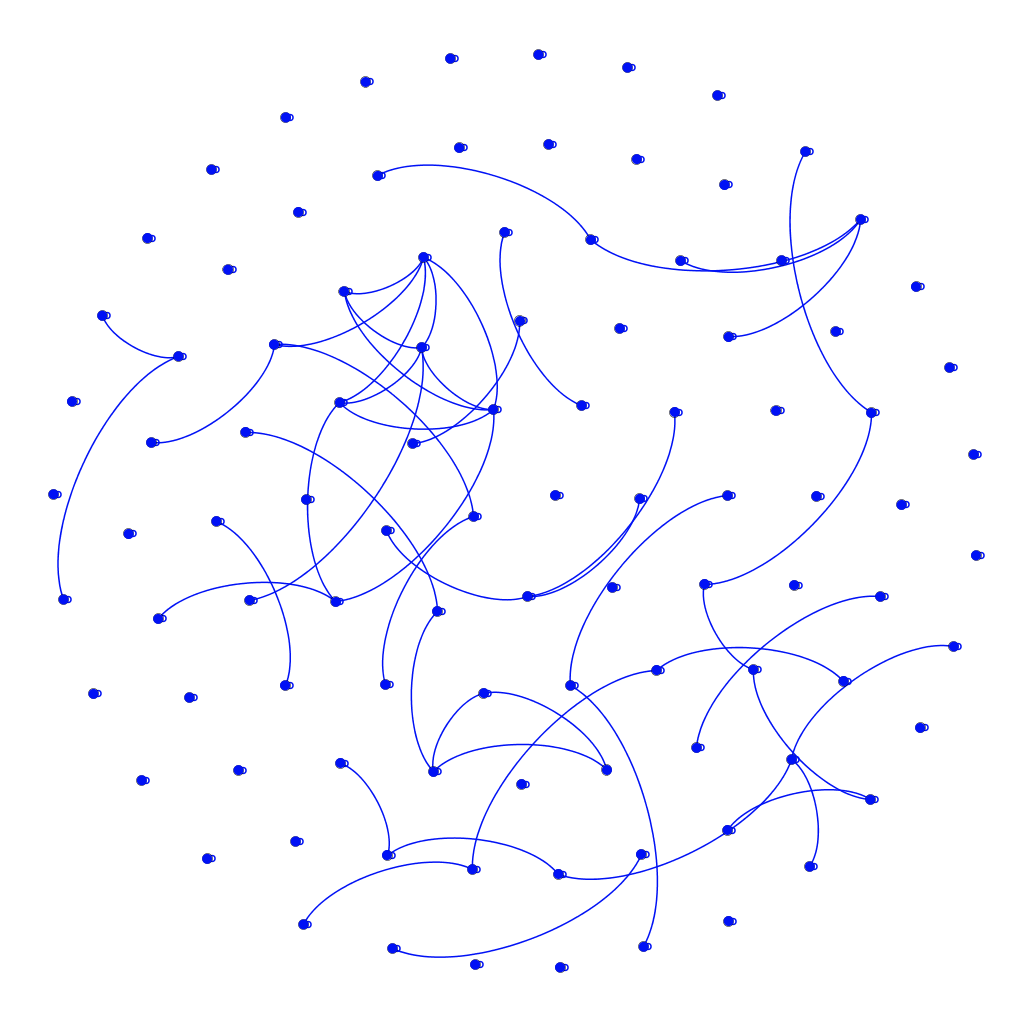}\\
(d) RR-debiased\\
\end{minipage}\vspace{-6pt}
\caption{Examples of differentially privately sanitized CTNs via RR and RR-debiased} \label{fig:visualnetworkS}\vspace{-9pt}
\end{figure}

\begin{figure}[!htb]
\centering\vspace{-3pt}
\includegraphics[width=1\textwidth]{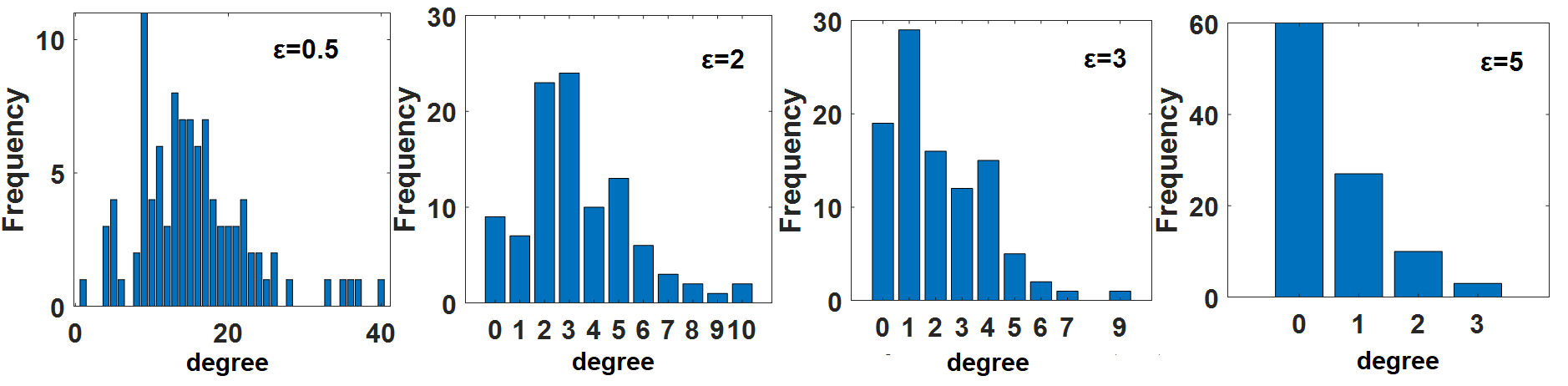}\\
(a) RR\vspace{-3pt}
\includegraphics[width=1\textwidth]{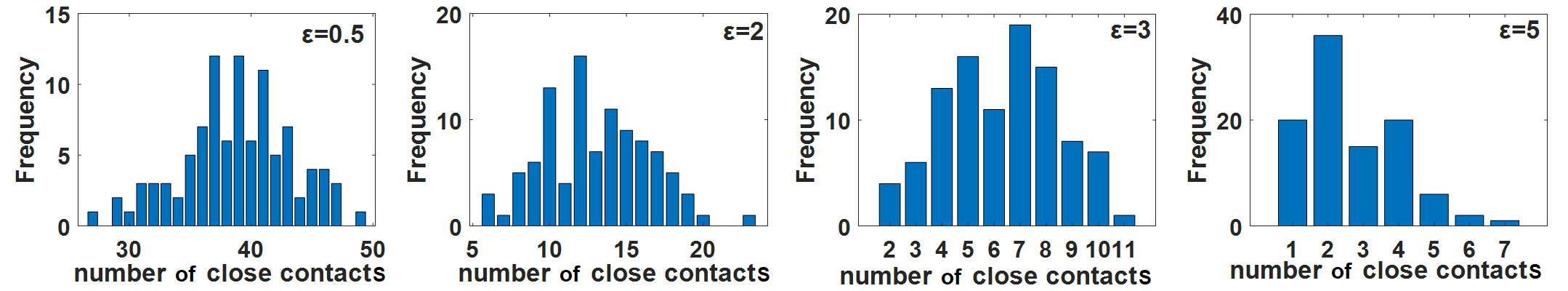}\\
 (b) RR-debiased\vspace{-3pt}
\caption{Degree distribution in the original and sanitized CTNs via RR and RR-debias}\label{fig:DDS}\vspace{-6pt}
\end{figure}

\begin{figure}[!htb]\centering
\includegraphics[width=0.375\linewidth]{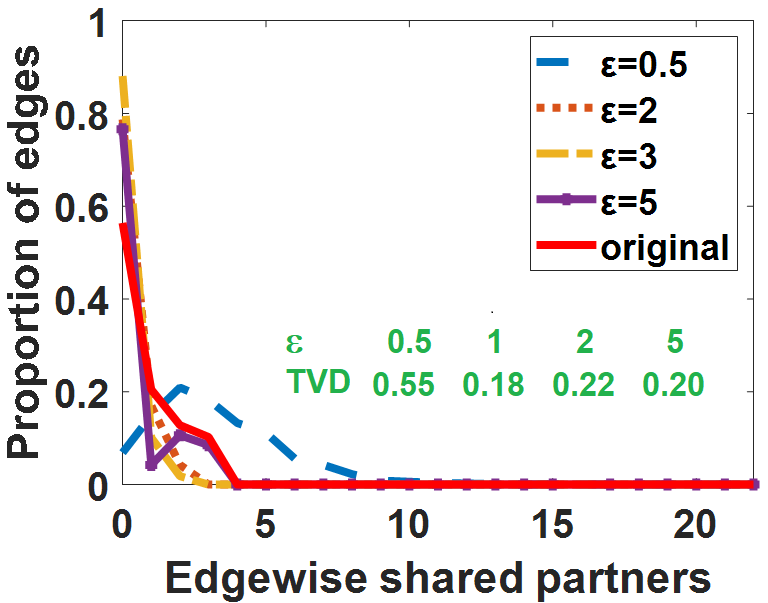}
\includegraphics[width=0.4\linewidth]{figure R1/ESPDPERM.png}\\
\hspace{0.2in} (a) RR \hspace{1.5in} (b) RR-debiased\\\vspace{-3pt}
\caption{Edgewise shared partner distribution in  sanitized CTNs via RR and RR-debias}\label{fig:ESPDGDDS}\vspace{-12pt}
\end{figure}

\begin{figure}[!htb]
\hspace{2cm} \textcolor{red}{betweenness centrality} \hspace{1in} \textcolor{red}{closeness centrality}\\
\includegraphics[width=0.245\linewidth]{figure R1/RRoribetweeness.png}
\includegraphics[width=0.245\linewidth]{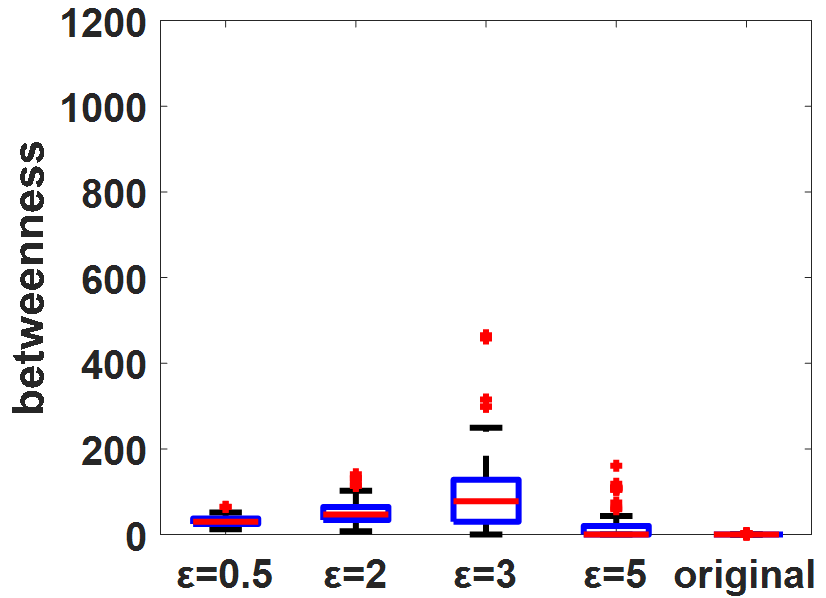}
\includegraphics[width=0.24\linewidth]{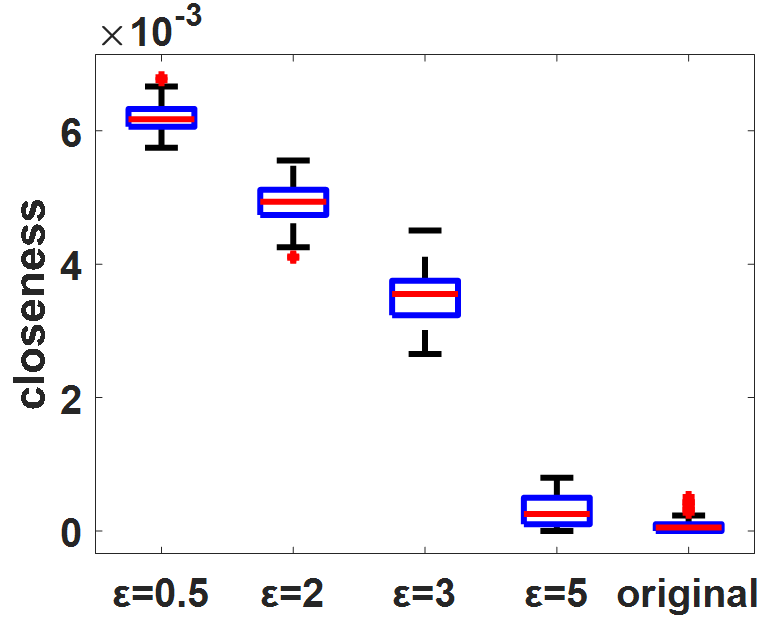}
\includegraphics[width=0.24\linewidth]{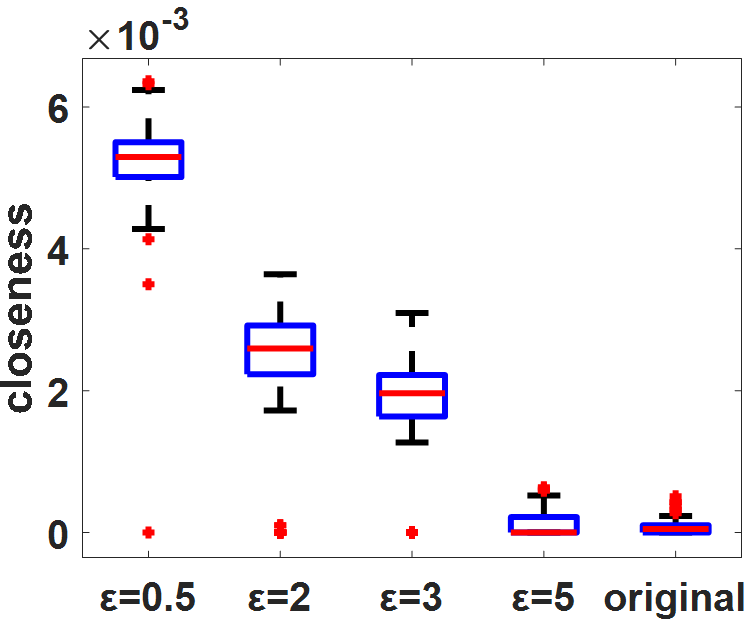}
$\mbox{\hspace{0.7in}}$ RR \hspace{0.5in} RR-debiased  \hspace{0.5in}  RR \hspace{0.5in} RR-debiased
\caption{Box plots of betweeness centrality and closeness centrality of 100 nodes in original and sanitized CTNs via RR and RR-debias} \label{fig:centralityS}\vspace{-12pt}
\end{figure}

The  sanitized  CTNs via RR and RR-debiased are presented in Figure \ref{fig:visualnetworkS} with the original CTN presented for comparison.  Table \ref{tab:edge.triangleS} presents the number of edges and number of triangles of the sanitized CTNs via RR and RR-debias. Figure \ref{fig:DDS} shows the DD, which is the distribution of close contacts of an individual in a CTN via RR and RR-debias and  Figures \ref{fig:ESPDGDDS}  depicts the ESPD of the sanitized CTNs with the  TVD in DD between the sanitized and original CTNs. Figure \ref{fig:centralityS} shows the box plots of the  betweenness centrality and closeness centrality of the 100 nodes in the sanitized CTNs via RR and RR-debias vs the original. 

\begin{table}[!htb]
\centering\vspace{-6pt}
\caption{Average (SD) number of edges and number of triangles over 100 repeats}\label{tab:edge.triangleS} \vspace{-6pt}
\resizebox{0.9\linewidth}{!}{
\begin{tabular}{c| cc|cc}
\hline
&  \multicolumn{2}{c|}{RR}  & \multicolumn{2}{c}{RR-debiased}  \\
\cline{2-5}
$\epsilon$ & number of edges & number of triangle & number of edges & number of triangle\\
\hline 
0.5& 1876 (37.4) & 8802 (528.3) & 844 (82.4) & 1258 (356.4)\\
2  & 619 (20.9) & 320 (38.1) & 182 (21.9) & 14 (6.1)  \\
5  & 72 (5.2) & 10 (0.9) &  48 (6.2) & 6 (2.5) \\
8 & 40 (1.3) & 10 (0.3) &  40 (1.4) & 9 (0.7) \\
10 & 39 (0.5) & 10 (0.1) & 39 (0.4) & 10 (0)\\
\hline
\end{tabular}}
\resizebox{0.8\linewidth}{!}{
\begin{tabular}{c}
original: number of  edges  = 39; number of  triangle  = 10.\hspace{1in}.\\
\hline
\end{tabular}}\vspace{-9pt}
\end{table}


The results on the privacy-preserving inference of the ERGM based on the sanitized CTNs via RR and RR-debias are presented in Table \ref{tab:ergmS}. We present the results for $\epsilon=5,15,18,24$; the results at $\epsilon<5$ are even worse. 
\begin{table}[!htb]\vspace{-6pt}
\caption{Privacy-preserving Inference of $\beta$ in the ERGM model based sanitized CTNs via RR and RR-debiased ($m=3$; 500 repeats)}\vspace{-3pt} \label{tab:ergmS}
\centering
\resizebox{0.6\linewidth}{!}{
\begin{tabular}{cc cccc}
\hline
method & metric & $\epsilon=5$ & $\epsilon=15$ & $\epsilon=18$ & $\epsilon=24$ \\
\hline
&bias  & 3.260 &  0.627 & 0.269 & 0.023\\
RR & RMSE  & 3.260 &  0.635 & 0.299 &0.165\\
&CP & 0 & 0.002 & 0.366 & 0.832\\
\hline
&bias  & 1.500  & 0.305 & 0.147 & 0.008\\
RR-debiased & RMSE  & 1.501  & 0.330 & 0.204 &0.165\\
&CP & 0  &0.366 &0.704 & 0.830 \\
\hline
\end{tabular}}
\resizebox{0.6\linewidth}{!}{
\begin{tabular}{l}
Original data: bias = -0.021, RMSE = 0.171, and CP = 0.942. \\
\hline
\end{tabular}}
\end{table}

\end{document}